\documentclass[reprint]{revtex4-1}
\usepackage[mathlines]{lineno}
\usepackage{lipsum}
\usepackage{blindtext}
\usepackage[export]{adjustbox}
\usepackage{graphicx,epsfig,epstopdf}
\usepackage{amsmath,amssymb}
\usepackage{esvect}
\usepackage{mathtools}
\usepackage{color}
\usepackage{subfigure}
\usepackage{array}
\usepackage{tabularx}
\usepackage{multirow}
\usepackage{booktabs}
\usepackage{float}
\usepackage{natbib}
\usepackage{url}
\newcolumntype{L}[1]{>{\raggedright\arraybackslash}p{#1}}
\newcolumntype{C}[1]{>{\centering\arraybackslash}p{#1}}
\newcolumntype{M}[1]{>{\centering\arraybackslash}m{#1}}
\allowdisplaybreaks

\begin{document}

\title{
	Coherent control of wave beams via unidirectional evanescent modes excitation}
	\author{Shuomin Zhong\textsuperscript{1*}}
	\thanks{Email: zhongshuomin@nbu.edu.cn, xuchen.wang@kit.edu}
	\author{Xuchen Wang\textsuperscript{2*}}
	\author{Sergei A. Tretyakov\textsuperscript{3}}
	\address{1. School of Information Science and Engineering, Ningbo University, Ningbo 315211, China 
 \\
2.  Institute of Nanotechnology, Karlsruhe Institute of Technology, 76131 Karlsruhe, Germany
 \\3. Department of Electronics and Nanoengineering, Aalto University, Finland}

	%
 

	\begin{abstract}
		Conventional coherent absorption occurs only when two incident beams exhibit mirror symmetry with respect to the absorbing surface, i.e., the two beams have the same incident angles, phases, and amplitudes. In this work, we propose a more general metasurface paradigm for coherent perfect absorption, with impinging waves from arbitrary asymmetric directions. By exploiting excitation of unidirectional evanescent waves, the output can be fixed at one reflection direction for any amplitude and phase of the control wave. We show theoretically and confirm experimentally that the relative amplitude of the reflected wave can be tuned continuously from zero to unity by changing the phase difference between the two beams, i.e. switching from coherent perfect absorption to full reflection. We hope that this work will open up promising possibilities for wave manipulation via evanescent waves engineering with applications in optical switches, one-side sensing, and  radar cross section control.
	\end{abstract}
		\maketitle
		\section{Introduction}	
		
Coherent control of propagation of a wave beam by tuning the amplitude and phase of another beam is a very promising approach to realize ultra fast optical devices for optical computing, sensing, and other applications \cite{fu2012all,2014Ultrafast,Shi:14,papaioannou2016two,papaioannou2016all,fang2015controlling,Silva160,SpatialProcessorCaloz,Zhu:20,kang2022coherent,peng2022coherent}. One of the most important effects in coherent control of light is coherent perfect absorption \cite{2010Coherent,2011Time,2012Coherent,2017Coherent,pirruccio2016coherent,Jung:15,Yoon:15,kita2017coherent,2018fibre,science.CPA,PhysRevB.91.220301}. In these devices, the level of absorption of one beam illuminating a thin sheet is controlled by another coherent beam that illuminates the same sheet. 

In earlier works, coherent perfect absorption (CPA) was achieved  only when with illumination from different sides of a homogeneous lossy layer and for two incident waves at the same angle \cite{2010Coherent,2011Time,PhysRevB.91.220301,2017Coherent}. The mechanism of coherent perfect absorption is destructive cancellation of all scattered beams. For homogeneous coherent perfect absorbers, there are  only specular reflection and non-diffractive transmission, allowing  coherent absorption  only with illumination of both sides and  at the same incidence angle. From the theoretical point of view and for many applications, it is important to  achieve coherent control of output for illuminations from the same side of the metasurface sheet at two or more arbitrary incidence angles. In Refs.~\cite{2012Measurement,Jung:15,Yoon:15}, coherent perfect absorption and scattering for two angularly asymmetric beams are realized by using surface plasmon-polariton (SPP) excitation at  silver-based diffraction groove gratings. However, such plasmonic grating designs have limitations. In particular, the structures are non-planar and operate only for TM modes at optical frequencies, where SPP are supported. Moreover,  there are always two output beams for different values of the phase of the control waves,  one of which may cause undesired noise to the useful  output signal due to parasitic scattering. This issue is critical in applications such as optical computing \cite{PhysRevApplied.11.054033}.

In this decade, the emergence of gradient metasurfaces \cite{yu2011light,2012Gradient,kildishev2013planar,epstein2016synthesis} and metagratings \cite{ra2017metagratings,epstein2017unveiling,popov2018controlling,wong2018perfect,cao2019mechanism,fu2019reversal,zhang2020coherent} has opened a new avenue for manipulation of light for arbitrary incidence angles and versatile functionalities. For periodical metasurfaces or metagratings with the period larger than half of the wavelength, the incident plane wave from one direction will be scattered into multiple directions, and the power carried by the incident wave can be redistributed among a number of diffraction modes. Based on this concept, several metasurface devices with perfect anomalous reflection working at microwaves \cite{sun2012high,D2017From} and optical bands \cite{he2022perfect} have been developed. However, in these previous works, the functionality of metasurfaces is designed only for one incident angle and the response for other illuminations is actually not considered. To design metasurfaces with coherent control functions for multiple simultaneously incident coherent beams from different directions, the matching conditions of amplitude, phase, and wavevector(direction) of the scattering modes between all incidences are required \cite{zhang2020coherent,cuesta2021coherent,cuesta2022coherent}, which is almost an impossible task using traditional gradient phase methods \cite{yu2011light,sun2012high} and  brute-force numerical optimizations \cite{D2017From,2018Extreme}. 

In this work, we perform inverse designs of CPA metasurfaces by solving the surface impedance satisfying the boundary condition determined by two coherent incident waves from two arbitrary angles and the desired total scattered waves. The engineering of evanescent waves in the scattered fields without altering the desired far-field outputs provides significant freedom in the CPA metasurface design, making another functionality of coherent control of reflection with  a single direction possible. It is demonstrated that excitation of unidirectional evanescent waves propagating along the surface in the direction of the incident-wave wavevector can be used to achieve single-direction output in coherently controlled optical devices. Furthermore, a mathematical optimization method based on scattered harmonics analysis \cite{PhysRevApplied.14.024089} is utilized to find the surface-impedance profile that simultaneously ensures the CPA and coherent maximum reflection (CMR) in a single direction. Thereafter, the substrate parameters are invoked as additional degrees of freedom in the optimization model, realizing reflection efficiency of 100\%. As an example, we experimentally validate the CPA gradient metasurface design in microwaves for TE-polarized waves by engineering the Indium Tin Oxide (ITO) film mounted on a grounded dielectric substrate. It is showed that the normalized output power can be continuously controlled between 0 and 1 by tuning the phase of the control wave.
	
\section{Design Concept}

	\begin{figure}[h!]
		\centering
		\includegraphics[width=0.9\linewidth]{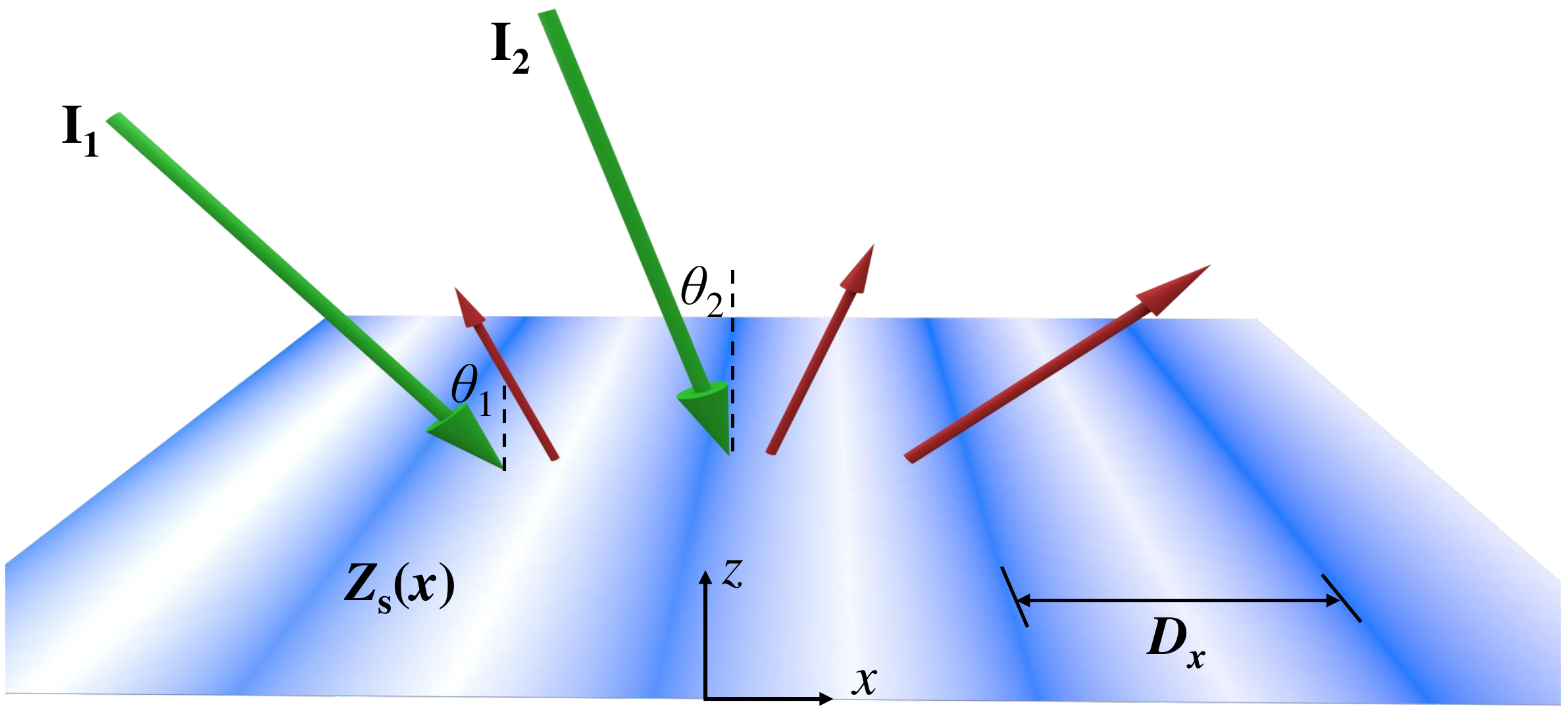}
		\caption{General scattering scenario for a periodically modulated impenetrable impedance surface. Two coherent  beams $I_1$ and $I_2$ are simultaneously incident from two angles.}\label{fig:Fig1}
	\end{figure}

Let us consider an impenetrable reciprocal metasurface whose surface is periodically modulated along the $x$-direction, with the period $D_x$. The surface is in the $xy$-plane of a Cartesian coordinate system (see Fig.~\ref{fig:Fig1}). The metasurface is simultaneously illuminated by two TE($s$)-polarized  plane waves $I_1$ and $I_2$ at the incidence angles $\theta_1$ and $\theta_2$  ($\theta_1>\theta_2$). The electric field amplitudes of the two beams $I_1$ and $I_2$ is $E_{\rm 1}=E_{\rm 0}$ and $E_{\rm 2}=\alpha E_{\rm 0}$, respectively ($\alpha$ is the amplitude ratio). The phase difference between them is $\Delta\phi$=0, defined at the origin point ($x=0,z=0$). The electromagnetic properties of the metasurface can be characterized by the locally-defined  surface impedance  that stands for the ratio of the tangential electric and magnetic field amplitudes at the surface plane $Z_{\rm s}(x)=E_{\rm t}(x)/H_{\rm t}(x)$.
		\begin{figure*}[bt!]
		\centering
		\subfigure[]{\includegraphics[width=0.43\linewidth]{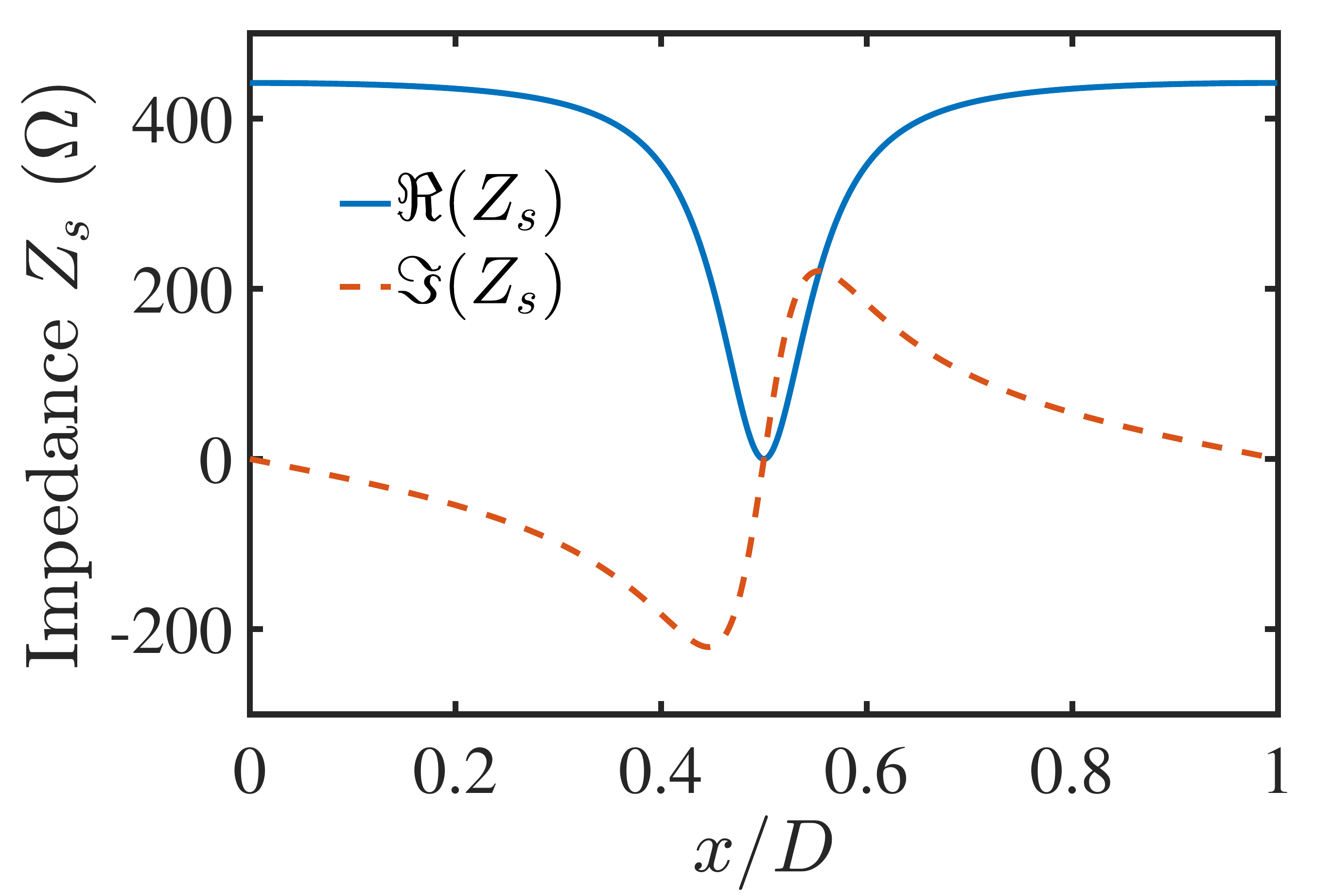}\label{fig:fig2a}}
		\subfigure[]{\includegraphics[width=0.45\linewidth]{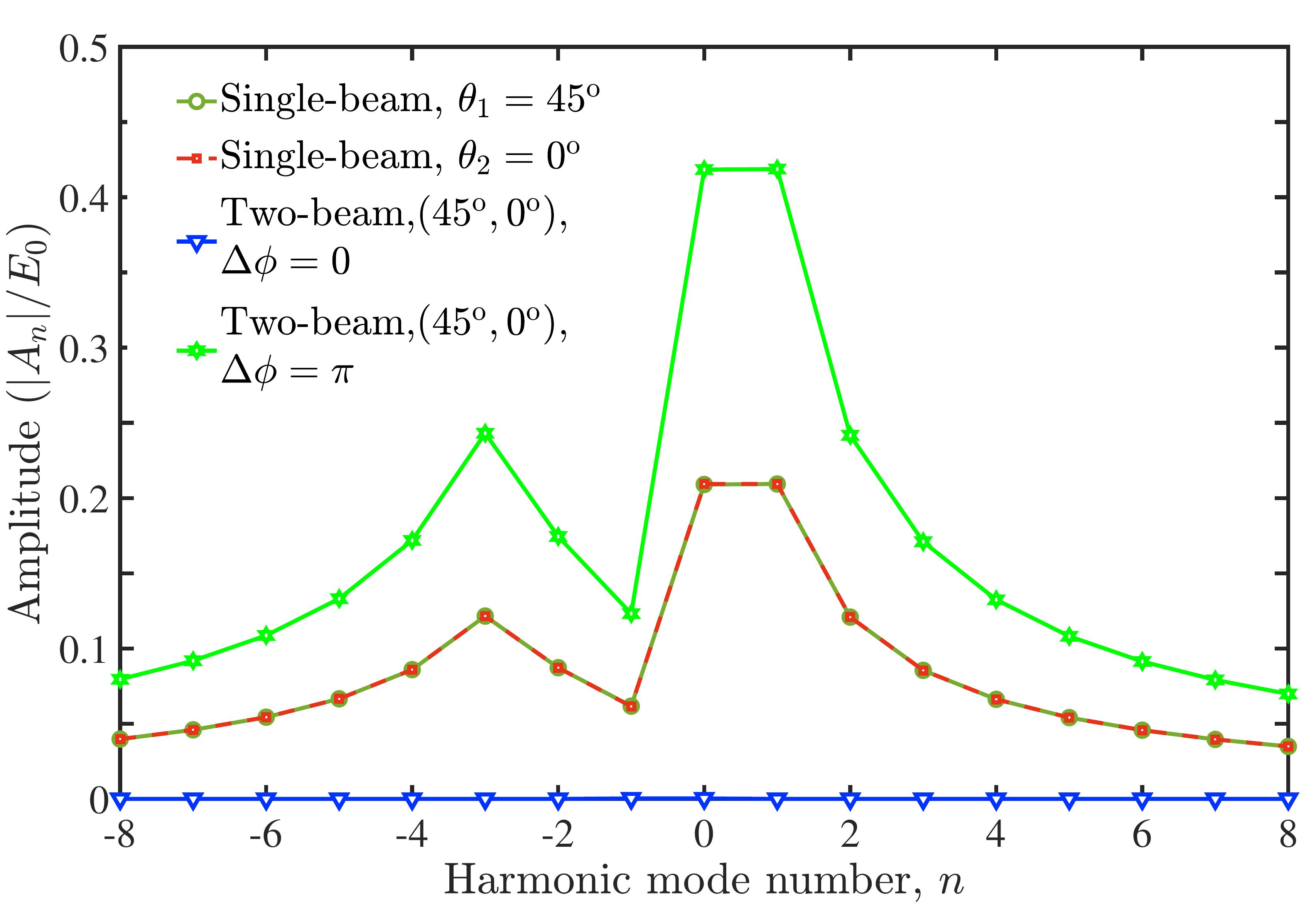}\label{fig:fig2b}}
		\subfigure[]{\includegraphics[width=0.43\linewidth]{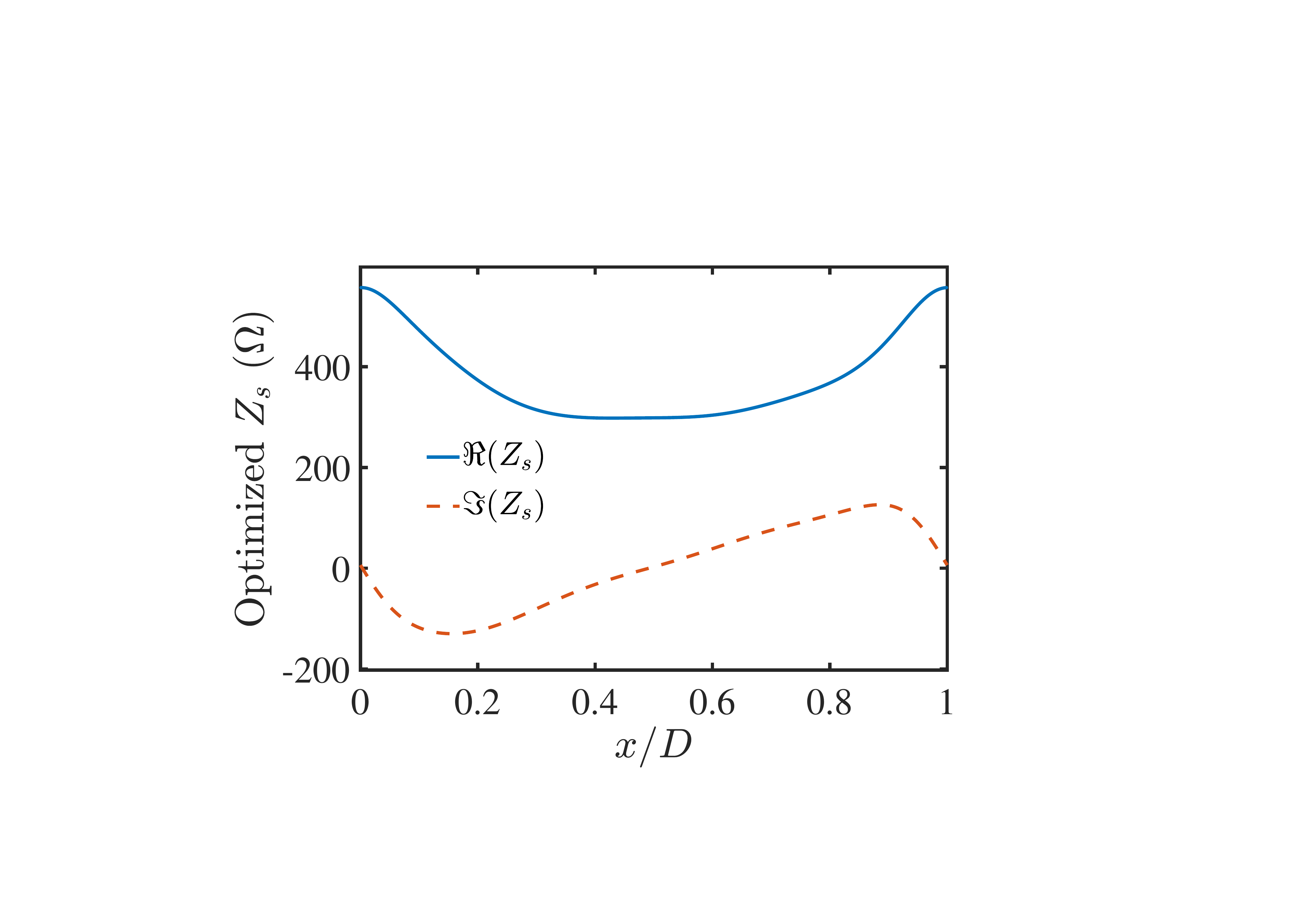}\label{fig:fig2c}}
		\subfigure[]{\includegraphics[width=0.45\linewidth]{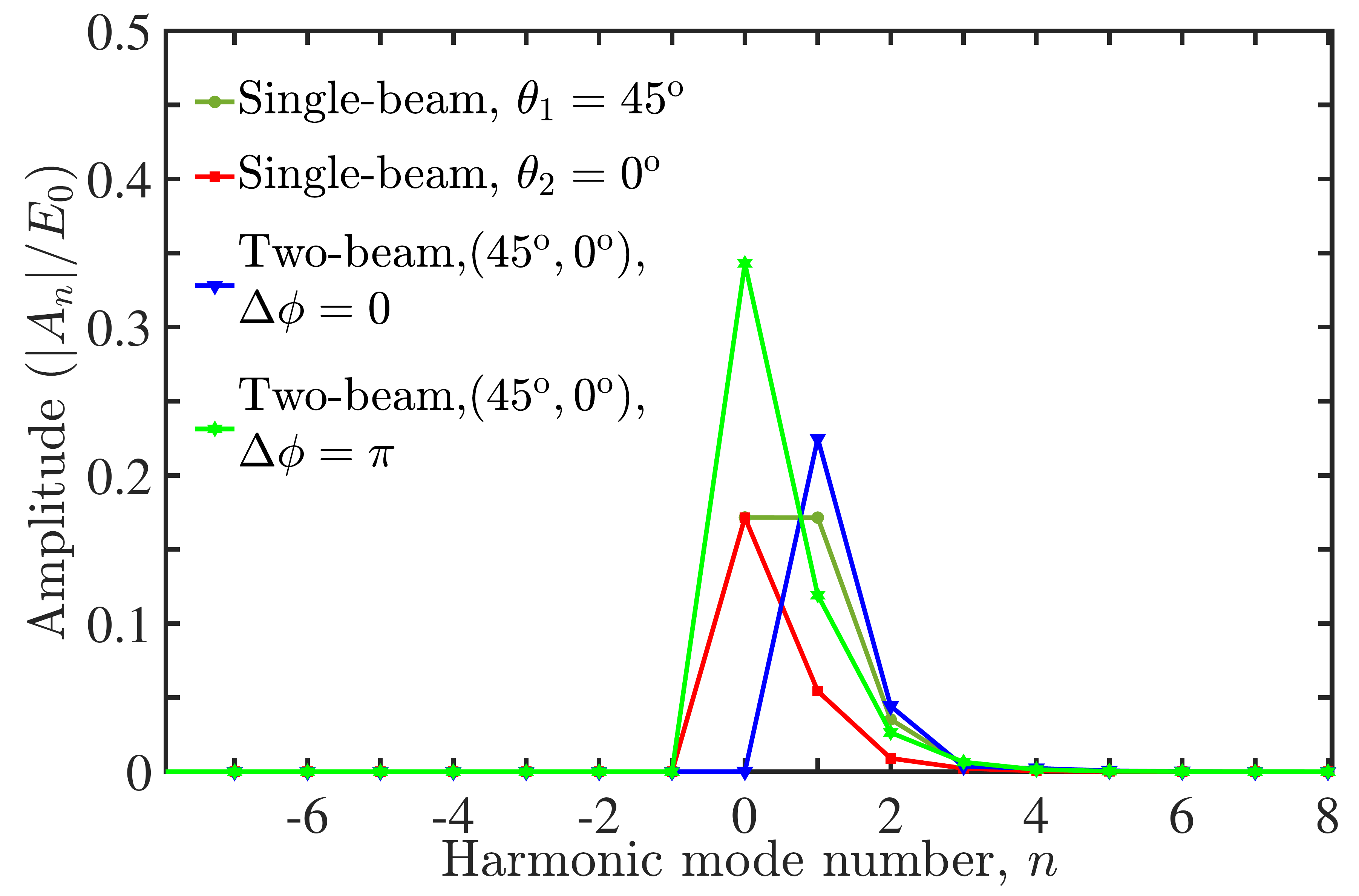}\label{fig:fig2d}}
	\caption{(a) Analytical surface impedance over one period to realize CPA for two incidence beams with $(\theta_{\rm 1},\theta_{\rm 2})=(45^\circ,0^\circ)$. (b) Magnitudes  of the complex amplitudes of different Floquet scattered harmonics (normalized by the amlpitude of the incident electric field $ E_0 $) when the gradient surface is illuminated by single-beam incidences at $ 45^\circ $ and $0^\circ $, and for two-beam incidences in phase and out of phase, respectively. (c) Optimized surface impedance profile over one period to realize CPA for in-phase incidences and single-direction reflection for out-of-phase incidences. The optimized Fourier coefficients of $Y_{\rm s}(x)$ read $g_0=2.654\times 10^{-3}+j1.724\times 10^{-11}$, $g_1=-7.770\times 10^{-4}-j1.045\times 10^{-10}$, $g_2=-(6.565+j4.581)\times 10^{-5}$, $g_3=-9.143\times 10^{-8}+j5.720\times 10^{-6}$, $g_4=(-1.644+j1.992)\times 10^{-5}$.  (d) Amplitudes of scattered harmonics when the optimized gradient surface in (c) is illuminated by single-beam incidences at $ 45^\circ $ and $0^\circ $, and for two-beam incidences in phase and out of phase, respectively.}  
	\end{figure*}
	
The field reflected by a periodically modulated metasurface  can be interpreted as a sum of Floquet harmonics. The tangential wavenumber of the $n$-th harmonic is related to the period and the incident wavenumber $k_{\rm 0}$   as $k_{{\rm r}xn}=k_{\rm 0}\sin\theta_i+{2\pi n_i}/{D_x}$, where $i=1, 2$. 
The corresponding normal component of the reflected wavenumber equals $k_{{\rm r}zn}=\sqrt{k_{\rm 0}^2-k_{{\rm r}xn}^2}$. If $|k_{{\rm r}xn}|$ is greater than the incident wave number, the wave is evanescent and  it does not contribute to the far field. For the harmonic wave satisfying $|k_{{\rm r}xn}|<k_{\rm 0}$, $k_{{\rm r}zn}$ is real, and this wave is propagating. The evanescent harmonics will be dissipated by the lossy surface and the propagating harmonics will propagate into the far-zone at the angles  $\theta_{rn}={\rm arcsin}(k_{{\rm r}xn}/k_0 )$. In order to achieve coherent perfect absorption, it is necessary (but not sufficient) to  ensure that all the diffracted propagating modes of two beams have the same set of angles $\theta_{rn}$, that allows mutual cancellation, defining the period  $D_x=\lambda_0/(\sin\theta_1-\sin\theta_2)$ \cite{See}, where $\lambda_0$ stands for the  wavelength.
	
Our aim is to achieve coherent perfect absorption for two coherent in-phase waves simultaneously incident on the metasurface at two different angles $\theta_1$ and $\theta_2$.  First, let us assume that no evanescent waves are excited  for these two illuminations. In the CPA case, there should be no reflected field at the surface. Thus, the tangential components of the total electric field at the plane $z=0$ can be written as $E_{\rm t}(x)=E_{\rm 0}(e^{-jk_0\sin\theta_{\rm 1}x}+\alpha e^{-jk_0\sin\theta_{\rm 2}x})$, where the time-harmonic dependency in the form $e^{j\omega t}$ is assumed and suppressed. 
The corresponding total magnetic field reads $H_{\rm t}(x)=E_{\rm 0}(\cos \theta_{\rm 1}e^{-jk_0\sin\theta_{\rm 1}x}+\alpha \cos \theta_{\rm 2}e^{-jk_0\sin\theta_{\rm 2}x})/Z_0$, with $Z_0=\sqrt{\mu_0/\epsilon_0}$ being the free-space wave impedance. The ratio of these electric and magnetic fields gives the required surface impedance  
	\begin{equation} 
	\newcommand\scalemath[2]{\scalebox{#1}{\mbox{\ensuremath{\displaystyle #2}}}}
	\begin{array}{c}\displaystyle
	\Re(Z_{\rm s})=\scalemath{0.95}{Z_0\frac{\cos\theta_{\rm 1}+{\alpha}^2\cos\theta_{\rm 2 }+{\alpha}\cos\Phi(\cos\theta_{\rm 1}+\cos\theta_{\rm 2 })}{\cos^2 \theta_{\rm 1}+{\alpha}^2\cos^2\theta_{\rm 2}+2{\alpha}\cos\theta_{\rm 1}\cos\theta_{\rm 2}\cos\Phi}},\\
	\\\displaystyle
	\Im(Z_{\rm s})=\scalemath{0.95}{Z_0\frac{{\alpha}(\cos\theta_{\rm 1}-\cos\theta_{\rm 2 })\sin\Phi}{\cos^2 \theta_{\rm 1}+{\alpha}^2\cos^2\theta_{\rm 2}+2{\alpha}\cos\theta_{\rm 1}\cos\theta_{\rm 2}\cos\Phi}}, 
	\end{array}
	\label{surface impedance}
	\end{equation}
where $\Phi=k_0(\sin\theta_{\rm 1}-\sin\theta_{\rm 2})x$ is the linearly varying phase. The real and imaginary parts of the surface impedance are even and odd functions of $x$, respectively. As is seen from Eqs.~\eqref{surface impedance}, the periodicity of the surface impedance is $D=\lambda_0/(\sin\theta_{\rm 1}-\sin\theta_{\rm 2})$, in accord with the above analysis. For passive metasurfaces, the real part of the surface impedance must be non-negative. Consequently, the amplitude ratio should satisfy $\alpha \geq 1$ or $\alpha \leq \cos\theta_{\rm 1}/\cos\theta_{\rm 2}$ to ensure passive solution for CPA by the surface. 

As an example, we consider two incident waves with incidence angles of $(\theta_{\rm 1},\theta_{\rm 2})=(45^\circ,0^\circ)$ and the same amplitude, assuming $\alpha=1$ for simplicity. (Other scenarios with $(\theta_{\rm 1},\theta_{\rm 2})=(60^\circ,-30^\circ),(75^\circ,15^\circ) $  are illustrated in the Supplemental Materials\cite{See}, corresponding to different surface impedance profiles.) As is shown in Fig.~\ref{fig:fig2a}, everywhere on the surface its resistance is  non-negative, demonstrating that passive gradient periodic surfaces can realize CPA for two asymmetric incident beams. 

To analyze the mechanism of CPA by the periodic impedance surface further, we can determine the amplitudes of all the Floquet scattered harmonics for general plane-wave illumination, using the method reported in \cite{PhysRevApplied.14.024089}. The total reflected field can be represented as an infinite sum of Floquet harmonic modes:
	\begin{equation}
	E_{\rm r}=\sum_{n=-\infty}^{\infty}A_ne^{-jk_{{\rm r}zn}z}e^{-jk_{{\rm r}xn}x}, 
	\label{second}
	\end{equation}
where $A_n$ is the  complex amplitude of the $n$-th Floquet harmonic.
Because the surface modulation is periodical, the surface admittance $Y_{\rm s}(x)=1/Z_{\rm s}(x)$ can be expanded into Fourier series:
	\begin{equation}
	Y_{\rm s}(x)=\sum_{n=-\infty}^{+\infty}g_{n} e^{-j2n\pi x/D}.
	\label{Fourier_series_grid_impedance}
	\end{equation}
A Toeplitz matrix ${\bf Y}_{\rm s}$ which we call the \emph{admittance matrix} is determined only by the Fourier coefficients of the modulation function and  filled with  ${\bf Y}_{\rm s}(r,c)=g_{r-c}$ at the $r$-th row and $c$-th column.
The reflection matrix is found as \cite{hwang2012periodic}
	\begin{equation}
	{\bf \Gamma}=\left({\bf Y}_0+{\bf Y}_{\rm s}\right)^{-1}({\bf Y}_0-{\bf Y}_{\rm s}),\label{Eq: reflection matrix}
	\end{equation}\label{Eq: reflection}where ${\bf Y}_0={\bf Z}_0^{-1}$ is a diagonal matrix with its main entry representing the admittance of each space harmonic, which is ${\bf Y}_{\rm 0}(n,n) $ =${k_{rzn}/{\omega_0\mu_0}}$. The amplitudes $ A_n $ of reflected harmonics for a given $ m $-th order Floquet harmonic of the incident wave can be calculated as $ A_n = {\bf \Gamma}(n,m)$. Note that $\bf \Gamma$ is a $(2N+1)\times(2N+1)$ square matrix and the columns and rows of $\bf \Gamma$ are indexed from $-N$ to $+N$. When the surface is illuminated by two waves  simultaneously, the amplitudes of all the Floquet harmonics are linear superpositions of all harmonics.

As is seen from Fig.~\ref{fig:fig2b}, when the two incident waves are in phase, all the harmonics have zero amplitude, meaning that CPA with no reflected fields occurs. However, when the two incident waves are out of phase, the reflected harmonics come out, including both propagating modes and evanescent ones, proving that the perfect absorption effect is phase-coherent, different from perfect absorption for two angles \cite{PerfectAbsorptionTwoAngles}. To understand the mechanism of CPA in the metasurface better, the harmonics of the  reflected field when single beams illuminate the surface separately are calculated. As shown in Fig.~\ref{fig:fig2b}, the complex amplitudes of every scattered harmonic are equal and $ 180^\circ $ out of phase (the phases are not shown here) for $ 45^\circ $ and $ 0^\circ $ incidences, resulting in destructive cancellation when the two beams illuminate simultaneously in phase. Here, the propagating harmonic of the order $ n=0 $ is defined at the specular direction of $\theta_{\rm 1}$ for both incidences. By properly designing the metasurface with the periodicity of $D=\lambda_0/(\sin\theta_{\rm 1}-\sin\theta_{\rm 2})$, three propagating modes corresponding to $ n=0,-1,-2 $ are created, and all the diffracted modes for both incidences have the same wave vectors, ensuing coherent interference for all corresponding harmonics. In the out-of-phase incidence case, the amplitudes of all the scattered harmonics double as compared to the single-beam case, as shown in Fig.~\ref{fig:fig2b}.
	
The analytical method to solve the surface impedance  boundaries used above is based on the objective to realize CPA with the amplitudes of both scattered propagating and evanescent harmonics being zero when two coherent beams illuminate the metasurface simultaneously. Indeed, the amplitudes of evanescent surface modes can be nonzero without breaking the CPA condition, because they do not radiate into the far zone and their power will be dissipated at the lossy surface. Thus, the solution of the surface impedance to achieve CPA is not unique if a certain set of evanescent waves with unknown complex amplitudes is excited. In addition to CPA, we invoke another functionality of coherent control of reflection with single direction, i.e. eliminating the unwanted outgoing beams at $ n=-1,-2 $ orders and keeping the $ n=0 $ order with the maximal amplitude, when the two coherent incident beams are out-of-phase. In this case, finding  the complex amplitudes of infinite numbers of evanescent modes for each incidence scenario is difficult or even impossible. Thus, instead of using the analytical method of calculating the surface impedance profile according to the total fields on the boundary, we apply a mathematical optimization algorithm described in Ref.~\cite{PhysRevApplied.14.024089} and based on the scattering matrix calculation to find a surface impedance profile that simultaneously ensures the coherent control capability for absorption and reflection of the surface. First, the metasurface is modelled as in  Eq.~\eqref{Fourier_series_grid_impedance}. To suppress propagating modes at the negative orders ($ n=-1,-2 $) and ensure that only the reflection channel at $ 45^\circ $ is open, the Fourier series of the surface admittance ${Y_{ s}(x)}$ are set to be unilateral as $Y_{\rm s}(x)=\sum_{n=0}^{4}g_{n} e^{-j2n\pi x/D}$ with non-negative-order series coefficients being nonzero (only five coefficients from $g_{\rm 0}$ to $g_{\rm 4}$  are used for improving optimization efficiency). This setting is reasonable because the unilateral surface admittance, making the admittance matrix ${\bf Y}_{\rm s}$ a lower triangular matrix, can lead to the reflection matrix $\bf \Gamma$ also being a lower triangular matrix, as is seen from Eq. (\ref{Eq: reflection matrix}). Consequently, the scattered modes contain only components of non-negative orders ($n \geq 0$). This effect highlights the role of unidirectional evanescent fields as a mechanism of suppressing propagating modes at the negative orders ($ n=-1,-2 $). Moreover, to ensure that the grid is a passive metasurface, we need to impose constraints $ \Re(Y_s)\geq 0 $, i.e., $ \Re(g_0)\geq |g_1|+|g_2|+|g_3|+|g_4| $. Secondly, the optimization goal is formulated as 6 objectives, including $ (|A_0|,|A_{-1}|,|A_{-2}|)=(0,0,0) $ for the in-phase scenario, and $ (|A_0|,|A_{-1}|,|A_{-2}|)=(A_{\rm 0max},0,0) $ for the out-of-phase scenario, where $ A_{\rm 0max} $ is the maximum magnitude of reflection in the out-of-phase case. In each trial of the optimization, an array of ${g_n}$ is assumed, and the value of all the objectives are calculated using Eq.(\ref{Eq: reflection matrix}). The sum of errors calculated for all the objectives is defined as a cost function \textsl{C}. By employing \textsl{MultiStart} and \textsl{fmincon} optimization algorithms,  the maximum magnitude of the out-of-phase reflection $ A_{\rm 0max}=0.34 $ is searched out, and the minimum value of \textsl{C} close to zero is achieved, meaning that the solutions of the impedance profile to realize the desired EM responses including CPA and single-direction-reflection are obtained.

Figure~\ref{fig:fig2c} shows a typical optimized solution of the surface impedance, which exhibits positive resistance everywhere along the metasurface. The calculated amplitudes of scattered harmonics for single-beam incidences at $45^\circ$ and $0^\circ$, and for two-beam incidences in phase and out of phase, for the impedance profile in Fig.~\ref{fig:fig2c}, are given in Fig.~\ref{fig:fig2d}, revealing the unilateral characteristic of scattering. We can see that the propagating components at $ n=-1,-2 $ orders are suppressed successfully by exciting the unidirectional evanescent wave. The only remaining propagating reflected channel is $ n=0 $ order at the outgoing angle of $45^\circ $. When two incoming beams are in phase, the reflected propagating harmonic ($n=0$) of each beam  cancel each other because they have the same amplitude and $ \pi $-reflection-phase difference. Distinct from the zero-amplitude of all the harmonics for the in-phase CPA scenario in Fig.~\ref{fig:fig2b}, the CPA in Fig.~\ref{fig:fig2d} occurs with non-zero-amplitude evanescent modes in the $n\geq 1$ orders. The amplitude of reflected electric field at $45^\circ $ ($n=0$) is doubled into $ A_{\rm 0max}=0.34 $ when two incoming beams are out of phase ($\Delta \phi= \pi$). We can conclude that the reflected power at $ 45^\circ $ can be continuously controlled by phase tuning of the control beam. When the two beams are out of phase, the reflected power normalized by the incident beam power at $ 45^\circ $ has the maximum reflection efficiency of 11.56 \%.
	
\section{Optimization and Practical design}

	\begin{figure}[!ht]
	\centering
	\includegraphics[width=0.9\linewidth]{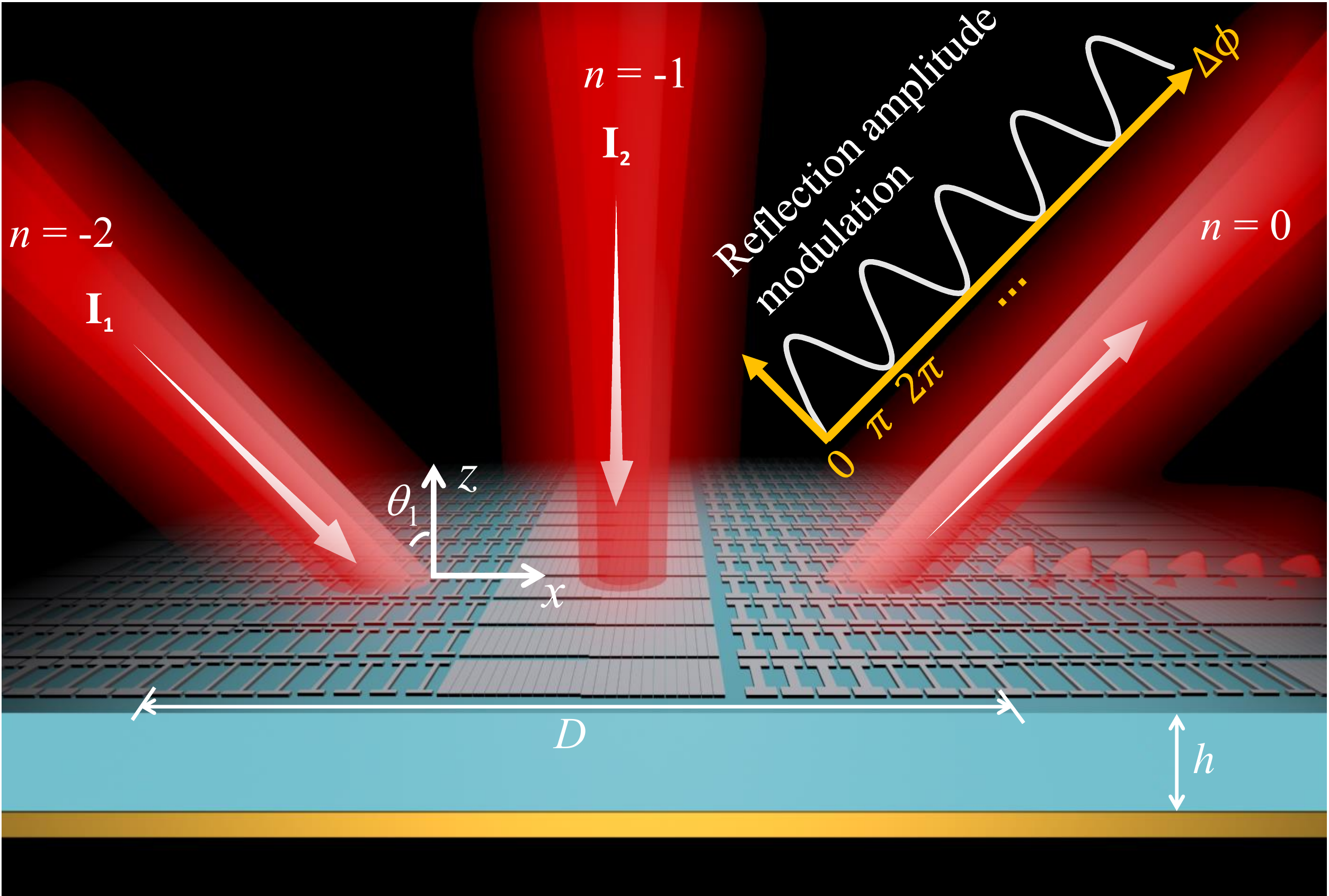}
	\caption{Schematics of reflection amplitude modulation for two coherent waves with the phase difference $\Delta\phi$ incident on a periodic sheet over a grounded dielectric slab. The amplitude of the output beam is modulated continuously by varying $\Delta\phi$, and switched between 0 (coherent perfect absorption) and 1 (coherent maximum reflection) when $\Delta\phi$ is switched between even and odd multiples of $\pi$. }\label{fig:fig3}
\end{figure}
	
 Low efficiency of the above design based on the impenetrable impedance model calls for optimization with the help of additional degrees of freedom. One possibility can be the use of one or more parameters of the actual implementation of the metasurface. 
	
In general, the impedance surface in the impenetrable model used above can be realized as a periodic metal pattern on a thin grounded dielectric slab, as shown in Fig.~\ref{fig:fig3}. The structure can be considered as a grid admittance of the top pattern with a shunt admittance of the grounded substrate. The characteristic admittance matrix $ {\bf Y}_{\rm d}  $ of the grounded substrate contains only diagonal terms $ {\bf Y}_{\rm d}(n,n) $, where ${\bf Y}_{\rm d}(n,n) $  is the admittance of the $ n $-th harmonic, and it is expressed as
\begin{equation}
{\bf Y}_{\rm d}(n,n)={k_{rzn}^{\rm d}}/[{j\mu_0\omega_0\tan(k_{rzn}^{\rm d}{h})}], \label{admittance}
\end{equation}
where $k_{rzn}^{\rm d}=\sqrt{\omega_0^2\epsilon_0\epsilon_{\rm d}\mu_0-k_{rxn}^2}$ is the normal component of the wavevector in the substrate (see Eq.S23 of the Supplemental Material of \cite{PhysRevApplied.14.024089}), $ \epsilon_{\rm d} $ and $ h $ are the permittivity and thickness of the substrate, respectively. The reflection matrix is calculated as
		$ {\bf \Gamma}=({\bf Y}_0+{\bf Y}_{g}+{\bf Y}_{\rm d})^{-1}({\bf Y}_0-{\bf Y}_{g}-{\bf Y}_{\rm d}) $.
When the thickness $ h $ is ultra-thin compared with the wavelength, for low-order harmonics we have $ \tan(k_{rzn}^{\rm d}h)\approx k_{rzn}^{\rm d}h$. As is seen from Eq.~(\ref{admittance}), the admittance for low-order harmonics equals approximately to $ 1/(j\mu_0\omega_0 h)  $, unrelated to the harmonic number. Thus, we can approximately design the top surface with the grid admittance $ Y_g(x)=1/Z_s(x)-{Y}_{\rm d}(0,0) $ using the optimized surface impedance $Z_s(x)$ in Fig.~\ref{fig:fig2c}, similar to Ref.~\cite{2018Extreme}. Due to the lack of freedom in the substrate design, the evanescent fields engineering is quite limited in the impenetrable model, resulting in a low reflection efficiency (11.56 \%) in the out-of-phase scenario. In order to implement CPA with a high reflection efficiency, we need to use the substrate parameters as additional degrees of freedom in the design. Since the admittance of the grounded substrate with a moderate thickness strongly depends on the harmonic number, the need of complicated matrix operations makes it impossible to analytically solve the grid impedance and substrate parameters. Thus, the optimization algorithm is extended by introducing the admittance matrix ${\bf Y}_{\rm d}$ of the grounded substrate, as described in Ref.~\cite{PhysRevApplied.14.024089}, to search for an optimum solution for the grid impedance profile and substrate thickness.
  		\begin{figure*}[bt!]
	\centering
	\subfigure[]{\includegraphics[width=0.44\linewidth]{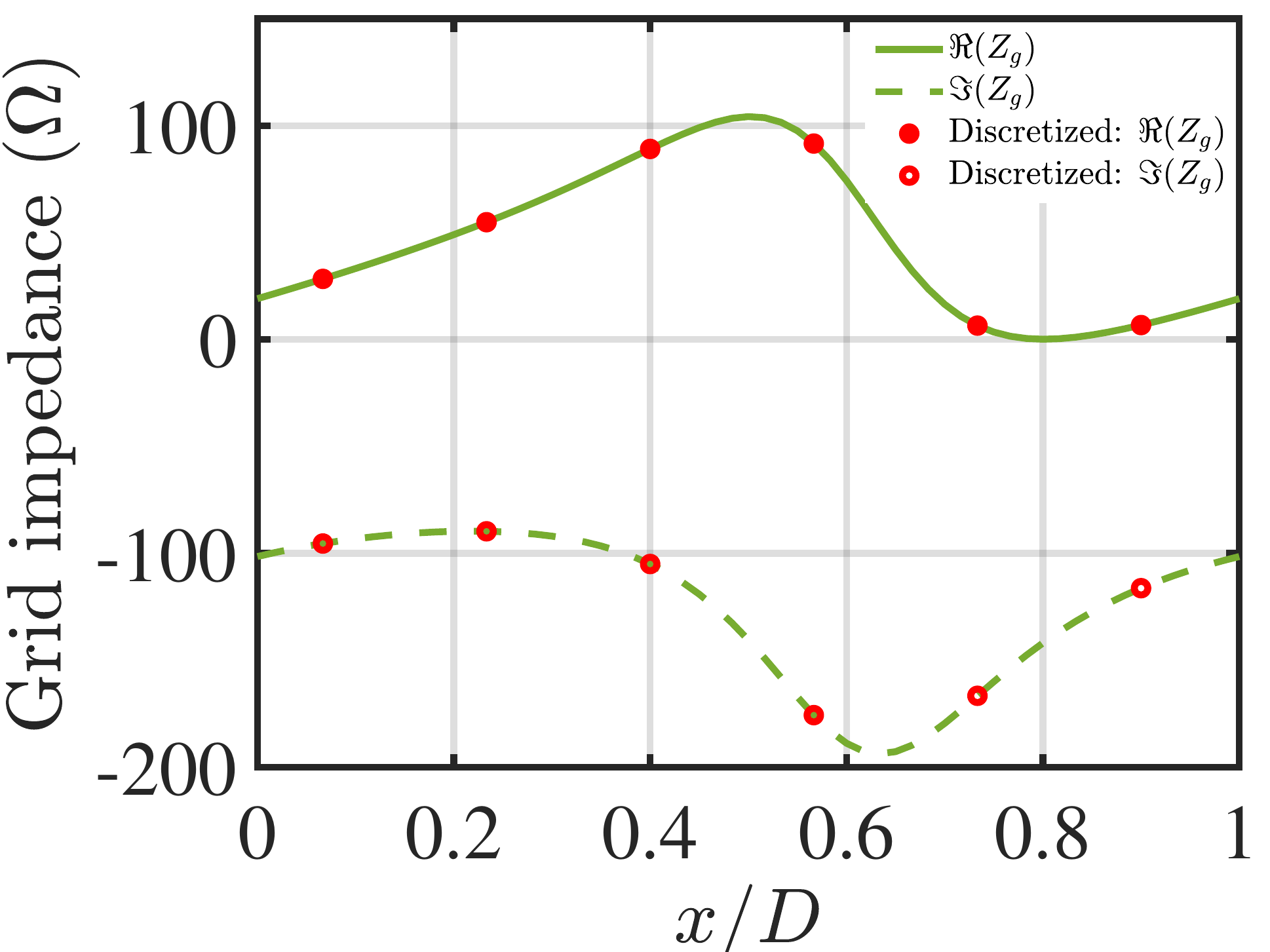}\label{fig:fig4a}}
	\subfigure[]{\includegraphics[width=0.46\linewidth]{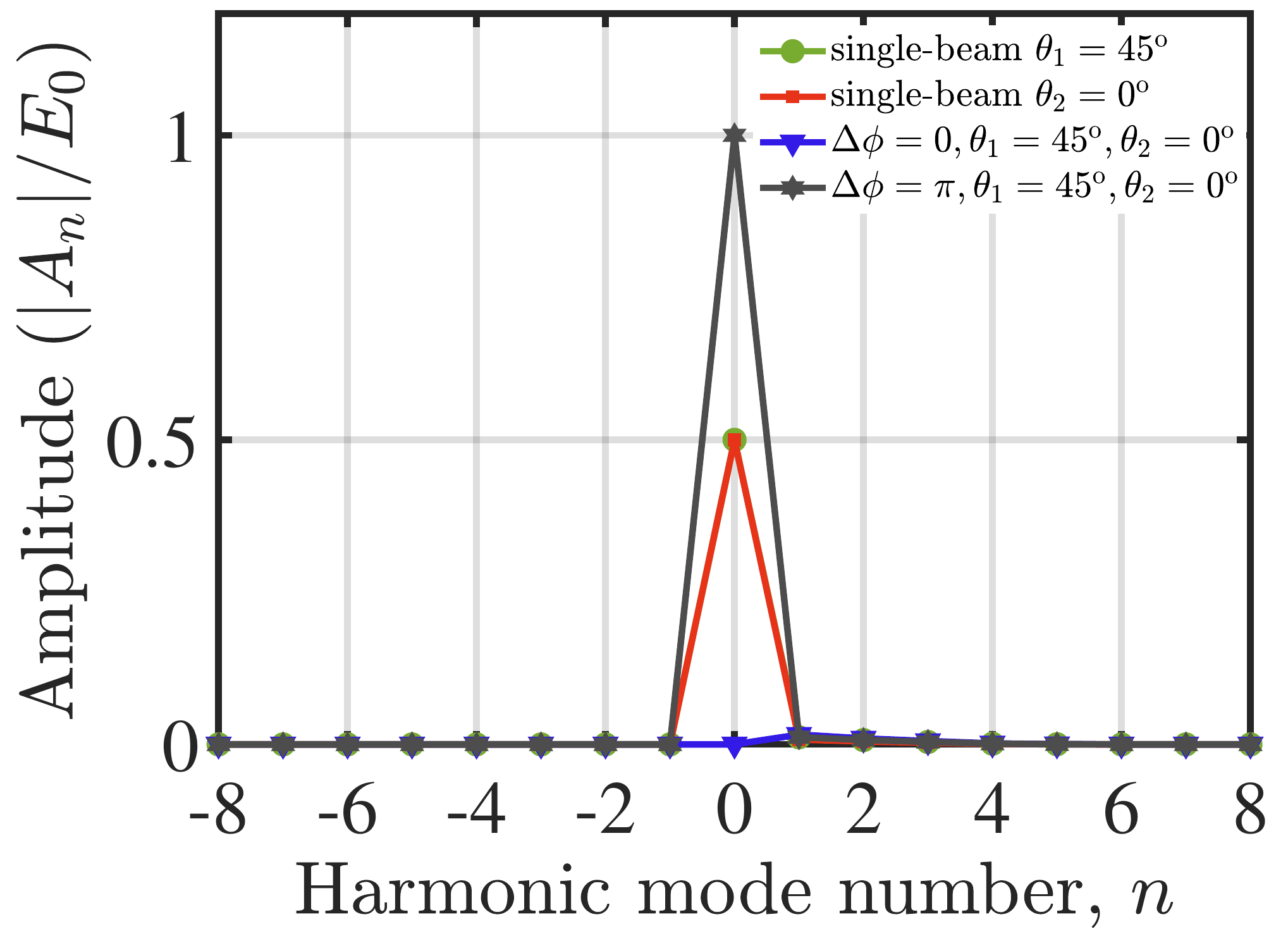}\label{fig:fig4b}}
	\subfigure[]{\includegraphics[width=0.33\linewidth]{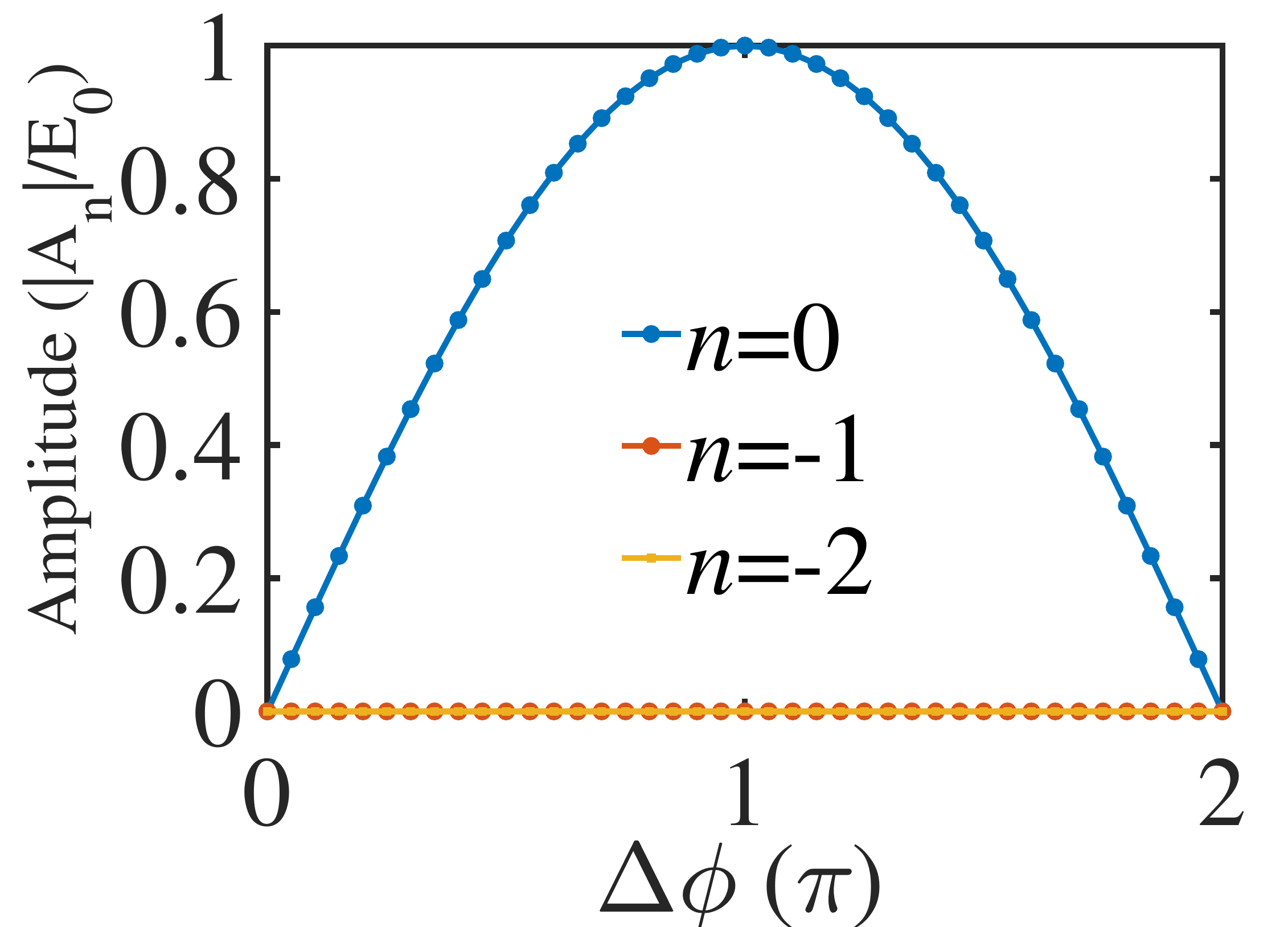}\label{fig:fig4c}}
	\subfigure[]{\includegraphics[width=0.53\linewidth]{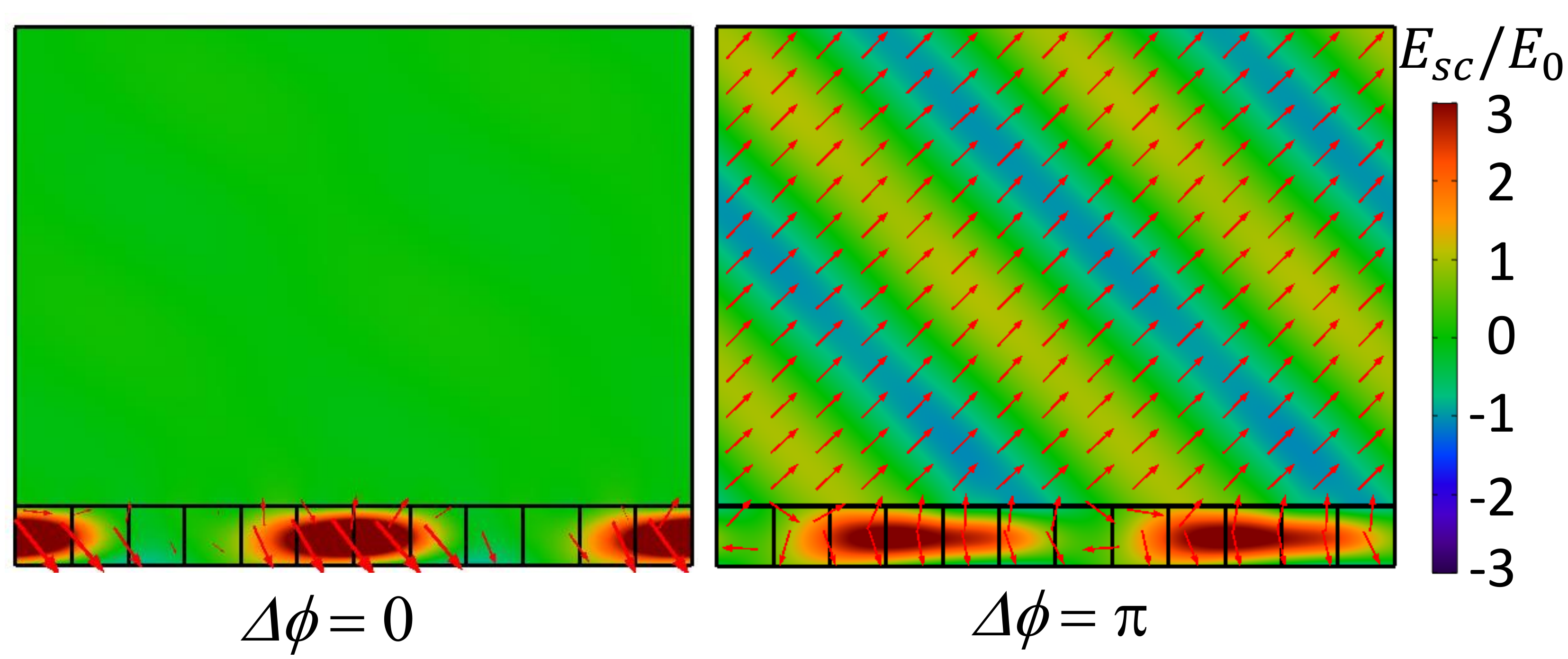}\label{fig:fig4d}}

		\caption{ 
		(a) The optimized and discretized grid impedance distribution over one period. (b) Amplitudes of the scattered harmonics when the optimized gradient metasurface is illuminated by a single beam  at $ 45^\circ $ and $0^\circ $, and for two-beam in-phase and out-of-phase illuminations, respectively. (c) The normalized amplitudes of three propagating harmonics ($ n=0,-1,-2 $) with a varying phase difference $\Delta\phi$ between incidences at $45^\circ$ and $0^\circ$. (d) The scattered electric fields and power density flow distributions  for the metasurface modeled by the discretized grid impedance (step-wise approximation, 6 subcells per period) on top of a grounded dielectric substrate. Two plane-wave incidences are in phase (left) and out of phase (right).}  
\end{figure*} 

 According to the results of the impenetrable model, the period of the impedance sheet modulation is set to $D=\lambda_0/\sin {45^\circ}$, with three propagating channels at $ -45^\circ $, $0^\circ$, and $ 45^\circ $. The Fourier series of the grid admittance is set to be unilateral as $ Y_{ g}(x)=g_{\rm 0}+g_{\rm 1} e^{-j2\pi x/D} $, ensuring that only the reflection channel at $ 45^\circ $ is open. In the optimization process, two Fourier terms $ g_{\rm 0}$ and $ g_{\rm 1}$ with four unknowns (the real and imaginary parts) are considered here to reduce complexity. The substrate thickness $ h $ is another unknown, and an available substrate with the permittivity  $ \epsilon_{\rm d}=5.8(1-j0.002)$ is used. The optimization goal is formulated as 6 objectives, the same as the objectives in the impenetrable model above. The constraints $ \Re(Y_g)\geq 0 $, i.e., $ \Re(g_0)\geq |g_1| $ are imposed to ensure the grid to be a passive metasurface. Additionally, to make the reactance easier to implement by patterning a thin conductive  surface,  another constraint $ \Im(g_0)\geq |g_1| $ is set to ensure that the surface reactance is always capacitive at all points of the metasurface.

The maximum magnitude of reflection $ A_{\rm 0max} $ in the out-of-phase scenario is searched out to be about 1 in the optimization, meaning that a reflection beam at $ 45^\circ $ with amplitude equal to the incident beam $I_1$ is obtained \cite{wang2021space}. It reveals that the invocation of substrate design provides an important additional  degree of freedom in engineering auxiliary evanescent modes to find a surface impedance that can realize the desired optimum scattering properties for all incidence scenarios. The optimized Fourier coefficients of the grid admittance $Y_{\rm g}(x)$ read $g_0=(2.599+7.054j)\times 10^{-3}$ and $g_1=(-0.807+2.463j)\times 10^{-3}$. The optimal substrate thickness is $ h = 0.2525\lambda_0$. The required grid impedance which is passive and capacitive along the metasurface is shown in Fig.~\ref{fig:fig4a}. 

Next, we analyse the scattered harmonics for the designed impedance sheet on the metal-backed dielectric substrate [see Fig.~\ref{fig:fig4b}]. The reflection coefficient of the metasurface has the same magnitude of 0.5 at $ n=0 $ order for $45^\circ$ and $0^\circ$ single-beam incidences, resulting from destructive interference when these two beams are in phase. For the out-of-phase scenario, the normalized magnitude of the reflected field at $ n=0 $ order ($45^\circ$) is about unity, which means that the reflected power efficiency reaches 100\% (normalized by the incoming power of the $45^\circ$ beam). Parasitic reflections into other directions ($ n=-1,-2 $) are seen to be negligible, due to the unilateral property of the admittance of the  surface. The evanescent harmonics are also unidirectional, but quite weak with the magnitude of 0.008 at $n=1$ order, and they are absorbed by the lossy structure, ensuring a CPA state. Figure~\ref{fig:fig4c} illustrates the phase-controlled modulation of reflections at three propagating orders. The reflection coefficient at $45^\circ$ can be continuously controlled from 0 to 1 by phase tuning, with the other two parasitic reflections maintained very close to zero. This phase-sensitive modulation between CPA and coherent maximum reflection (CMR) without parasitic reflections is important in light switching applications where a low-return-loss characteristic is required. See the Supplemental Animation \cite{See} for the switch of reflected beam by an incident phase-controlled wave.

In implementations, the influence of discretization on the metasurface performance is an important factor (see detailed analysis of scattered harmonics versus the number of subcells in Ref.~\cite{See}). We use six subcells over a period and each discretized impedance value is set at the central point of each subcell, as shown in Fig.~\ref{fig:fig4a}. The scattered fields from the ideal impedance sheet on the metal-backed dielectric slab for both in-phase and out-of-phase incidences are presented in Fig.~\ref{fig:fig4d}, using full-wave simulations in Comsol. The reflected field distribution confirms that the metasurface with six subcells per period possesses the desired response: nearly perfect absorption with reflection amplitude of only 0.023 for two in-phase illuminations and nearly total reflection at $45^\circ$  for two out-of-phase illuminations, relative to the intensity of the $45^\circ$ incidence. It is seen that the top lossy sheet and reflective ground separated by the slab act as a leaky-wave cavity with enhanced fields. For the in-phase scenario, the direct reflections of the top surface and leaky wave components of the cavity destructively cancel out, and all the power is absorbed by the lossy surface, causing CPA. By changing the initial phase difference between the two coherent incidences into $\pi $, constructive interference occurs among these components, which results in nearly total reflection. Note that in the out-of-phase case a half of the total incoming power (two incident beams) is still absorbed by the lossy surface.

\section{Physical implementation and experimental validation}	 
	
		\begin{figure*}[bt!]
		\centering

		\subfigure[]{\includegraphics[width=0.45\linewidth]{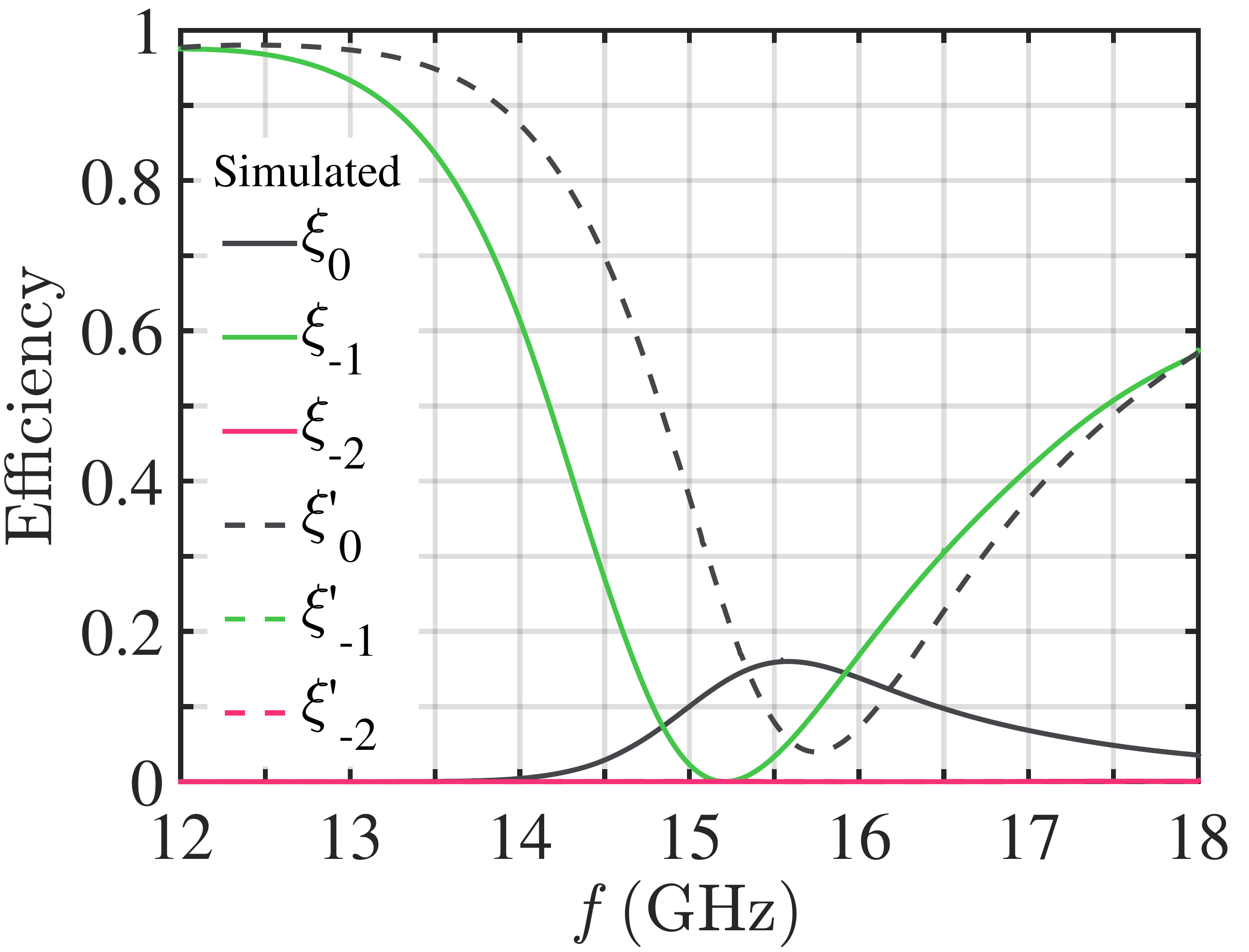}\label{fig:fig5a}}
	   	   \subfigure[]{\includegraphics[width=0.45\linewidth]{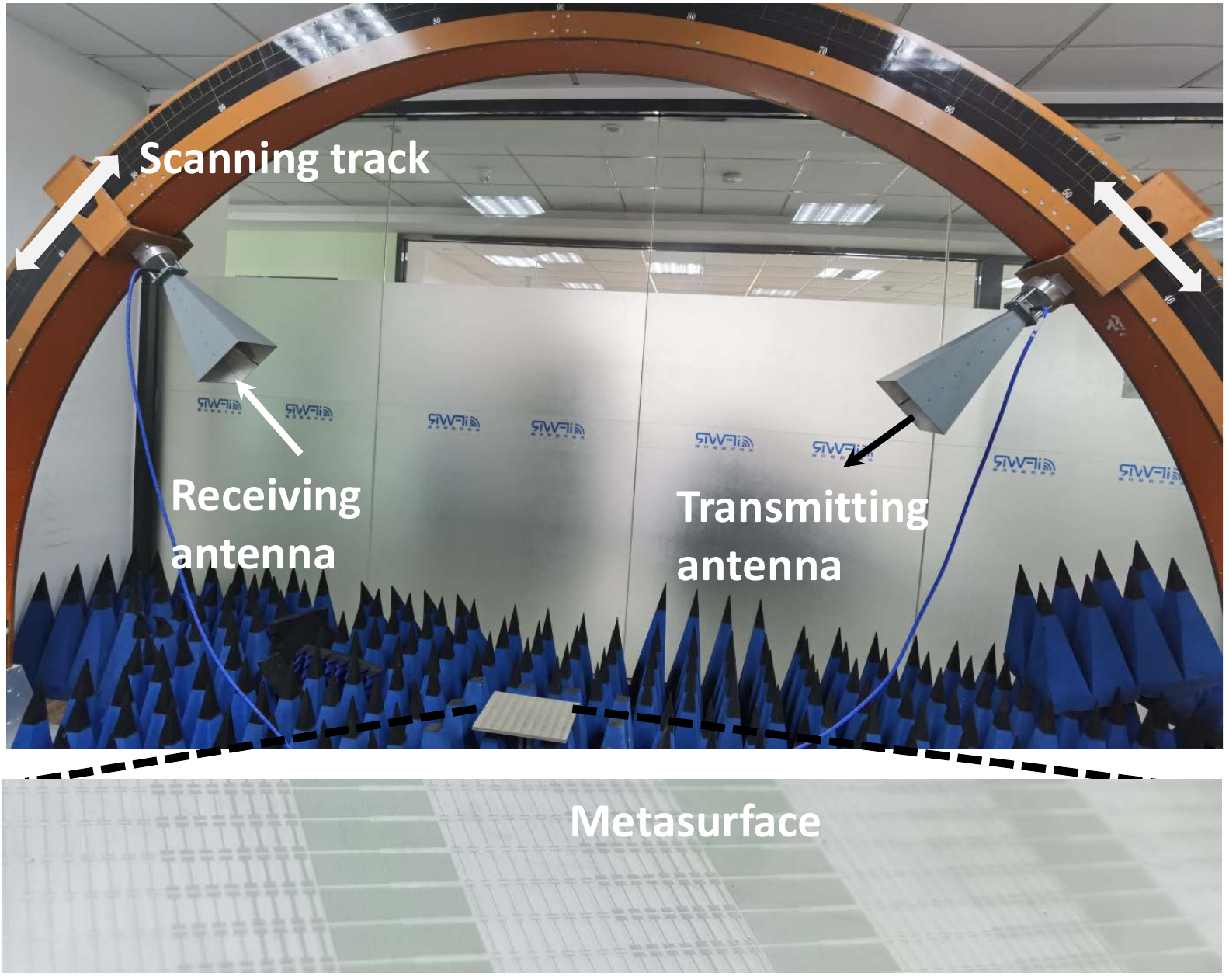}\label{fig:fig5b}}
	   	    \subfigure[]{\includegraphics[width=0.45\linewidth]{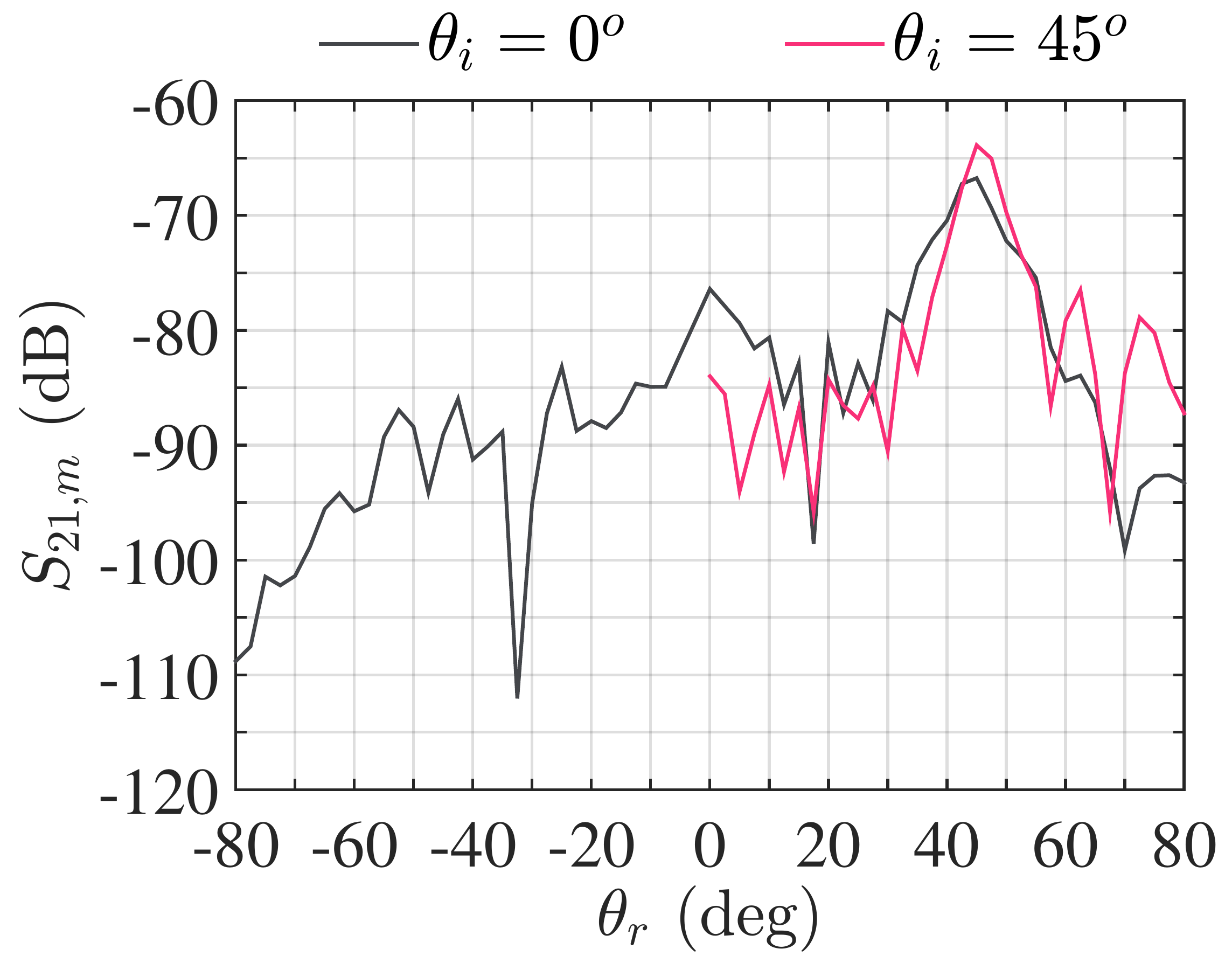}\label{fig:fig5c}}
	    \subfigure[]{\includegraphics[width=0.45\linewidth]{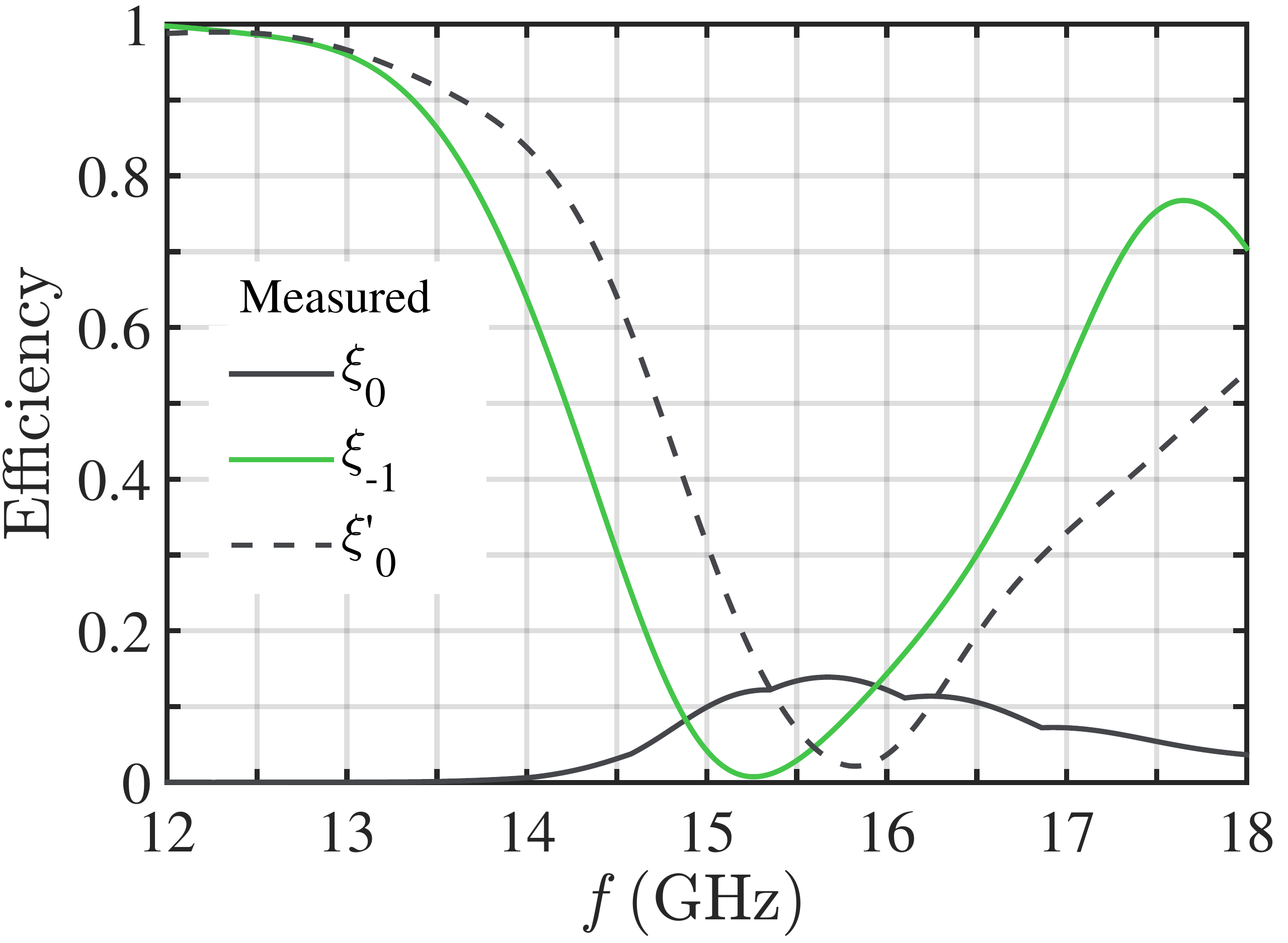}\label{fig:fig5d}}
				\caption{ 
			(a) Simulated and (d) measured reflection efficiency spectrum for different diffracted modes of each single beam at $ 0^\circ $ (solid lines) and $ 45^\circ $ (dashed lines). (b) Schematic of the experimental setup (top) and photograph of the fabricated sample (bottom). (c) Signals at 15.22~GHz measured by the receiving antenna at different orientation angles with the transmitting antenna at $ 0^\circ $ and $ 45^\circ $.} 
	\end{figure*}
	
The theory above is general and applies to any frequency, and we choose the microwave band for a proof of concept demonstration. The required impedance profile at 15.22 GHz is realized using an ITO film with the surface resistance of 5.5 $\Omega$/sq supported by a grounded dielectric slab with the thickness $h= 4.95$ mm, as shown in Fig.~\ref{fig:fig3}. The detailed parameters and structures of each unit cell are presented in the Supplementary Material\cite{See}. Due to the resolution limitation of picosecond laser micro-processing, the complex grid impedance is implemented as six subcells, and each subcell is divided into four equal sub-subcells in order to make the local design of the gradient impedance more robust. By structuring the homogeneous resistive ITO film into I-shaped cells, the required grid resistance and reactance on a surface in Fig.~\ref{fig:fig4a} can be created. For $y$-polarization incident waves, such I-shaped resonators can be modeled as RLC series circuits. The required resistance is realized by tailoring the width and length of the ITO strips. Smaller width and longer length result in higher grid resistance. The required reactance can be tailored by adjusting capacitance of the gap, which can be increased by narrowing the gap or increasing the length or width of the bar, with a small influence on the resistive part. The 5th and 6th subcells degenerate into strips, to implement resistive parts as close to the theoretical value as possible. However, there are still deviations of 3.6~$\Omega $ and 1.1~$\Omega $ from the theoretical resistances of the 5th and 6th subcells, respectively. The deviation can be eliminated if an ITO film with a lower surface resistance is utilized. To simplify the fabrication process, we neglect this deviation. The impact is analyzed theoretically, showing that the reflection amplitude in the in-phase scenario increases from 0.023 to 0.065, which is tolerable in experiments. Since the two beams with  $0^\circ $ and  $45^\circ $ incidence angles illuminate the surface simultaneously, all the elements should have angle-independent surface impedances. The I-shaped resonators have angle-insensitive impedance under TE incidences, satisfying this requirement  \cite{2008Simple}. In the strips of the 5th and 6th subcells, narrow slits are cut out to reduce the angular sensitivity of the impedance.  All the subcells have been optimized with the geometrical dimensions specified in Ref.~\cite{See}. 

 Figure~\ref{fig:fig5a} shows the simulated frequency response of the metasurface for the normal and $45^\circ $ incidences. For the normal illumination, strong  reflections occur at $n=-1$ and $n=0$ harmonics (denoted as $\xi_{-1}$ and $\xi_0$), and the amplitude of the $n=-2$ scattered propagating mode is nearly zero in the whole frequency band. The reflection at the $n=-1$ mode (specular reflection at $0^\circ $) also has a near-zero dip at the design frequency of $15.22$ GHz, and the reflection efficiency at the $n=0$ mode(anomalous reflection at $0^\circ $) is about $13.9 \%$ (the relative amplitude is 0.44). Note that for anomalous reflection, the efficiency is calculated as $\xi={(E_r/E_i)^2}{\cos\theta_r}/{\cos\theta_i}$ \cite{D2017From}. For the $45^\circ$ illumination, the reflections at both  $n=-1$ and  $n=-2$ modes ($\xi'_{-1}$ and $\xi'_{-2}$) are close to zero, and the efficiency at the $n=0$ mode ($\xi'_{0}$) is about $21 \%$ at $15.22$ GHz (the relative amplitude is 0.46). Therefore, at the operating frequency $15.22$ GHz, the reflected modes for both incidences at the outgoing angle of $45^\circ$ are almost equal-amplitude, satisfying the condition of CPA. The scattered electric field distributions of the designed metasurface illuminated by two beams in the in-phase and out-of-phase scenarios obtained from full-wave simulations are presented in Ref.~\cite{See}. It can be seen that when the two illuminations are in phase, the total scattered fields are quite small (0.02), indicating nearly perfect coherent absorption. However, when the two illuminations are switched into the out-of-phase state, the relative amplitude of the scattered fields is about 0.91, and the coherent maximum reflection is mainly along the $45^\circ $ direction.
    
We have fabricated a sample (see Methods) and carried out several experiments to validate the theoretical results (see Fig.~\ref{fig:fig5b}). First, the transmitting antenna is fixed at $0^\circ$, whereas the receiving antenna is moved along the scanning track with a step of $2.5^\circ $. The signal reflected from the metasurface is measured by the receiving antenna at different angles $\theta_r $. Then, the transmitting antenna is fixed at $45^\circ$ and the receiving antenna is scanning its position to measure the reflected signal in the other half space. As shown in Fig.~\ref{fig:fig5c}, the main peaks of reflections for both two incidences occur at $\theta_r=45^\circ $, which is an expected result according to the theory and simulations. There is another reflection peak at $\theta_r=0^\circ $  for the normal incidence case, which is about $-10$~dB lower than the main peak, corresponding to a low specular reflection at $15.22$ GHz. 

To estimate the amplitude efficiency of the metasurface at all three reflection channels, we replaced the metasurface by a copper plate of the identical size and measured the specular reflection signal amplitudes from the reference uniform metal mirror for $\theta_i=2.5^\circ $ (approximately normal incidence), $22.5^\circ $, and $45^\circ $ incidence angles. The specular reflection efficiency of the metasurface for $ 0^\circ $ and $ 45^\circ $ illuminations are calculated by normalizing the signal amplitude by the amplitude of the signal reflected from the reference plate, illuminated at $2.5^\circ$ and $45^\circ$ angles, respectively. As shown in Fig.~\ref{fig:fig5d}, at the design frequency of $15.22$~GHz, the specular reflection efficiencies at $0^\circ $ and $ 45^\circ $ ($\xi_{-1}$ and $\xi'_{0}$) equal $0.8 \%$ and $18.6 \%$ (the relative amplitude is 0.431), respectively. For the anomalous reflection at the $n=0$ mode for the normal incidence, the reflection angle is  $\theta_r=\arcsin(15.22/(\sqrt{2}f))$, which equals  $45^\circ $  at $15.22$~GHz and varies from $63.7^\circ $ to $36.7^\circ $  as the frequency changes from $12$~GHz to $18$~GHz. Therefore, we choose the signal data of a different receiving angle $\theta_r$ calculated according to different frequency band and normalize its signal amplitude by the signal amplitude from the reference mirror for different $\theta_r/2$ incidence angles. Additionally, we divide the obtained value by an estimated correction factor \cite{D2017From} $\sqrt{\cos(\theta_r)}/\cos(\theta_r/2) $, which gives the ratio between the theoretically calculated signal amplitudes from an ideal metasurface (of the same size and made of lossless materials) and a perfectly conducting plate. At the design frequency of $15.22$ GHz, the correction factor is equal to $0.91$, thus the reflection efficiency is calculated as $12 \%$(the relative amplitude is 0.412), as shown in Fig.~\ref{fig:fig5d}. The measured efficiency is in good agreement with the results obtained using numerical simulations (see Fig.~\ref{fig:fig5a}), except for some ripples in the $\xi_{0}$ curve caused by the discrete angular scanning step in the measurement. The relative amplitudes of reflections for both incidences at the $n=0$ mode are almost equal in the measurements, verifying the capability for CPA.
  
  \begin{figure*}[bt!]
		\centering
		\begin{minipage}[b]{0.5\textwidth}
		\centering
			\subfigure[]{\includegraphics[height=8cm,width=1\linewidth]{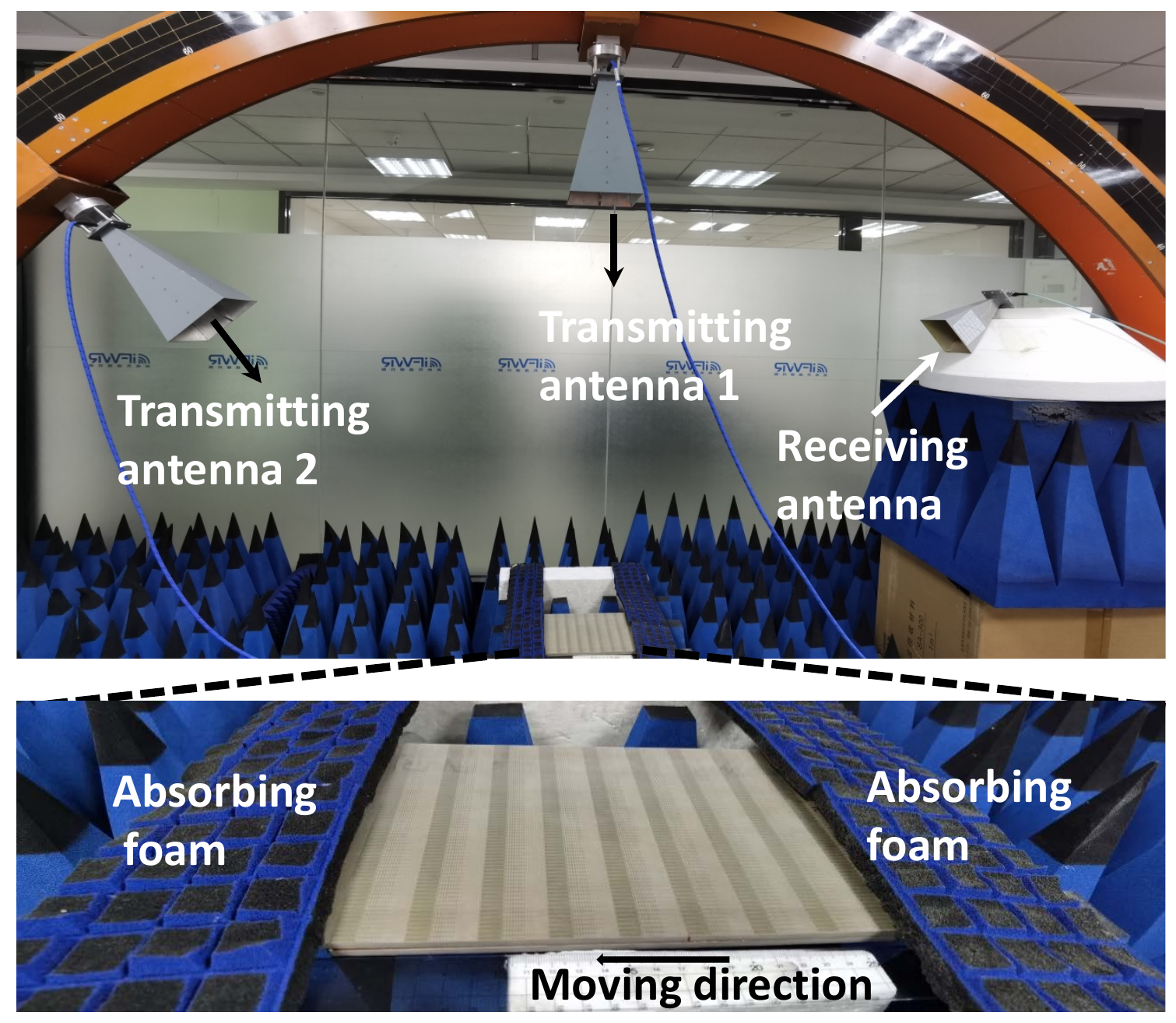}\label{fig:fig6a}}
		\end{minipage} 	
				\hspace{0.1cm}
			\begin{minipage}[b]{0.4\textwidth}
			\centering
	   	    \subfigure[]{\includegraphics[height=3.2cm,width=1\linewidth]{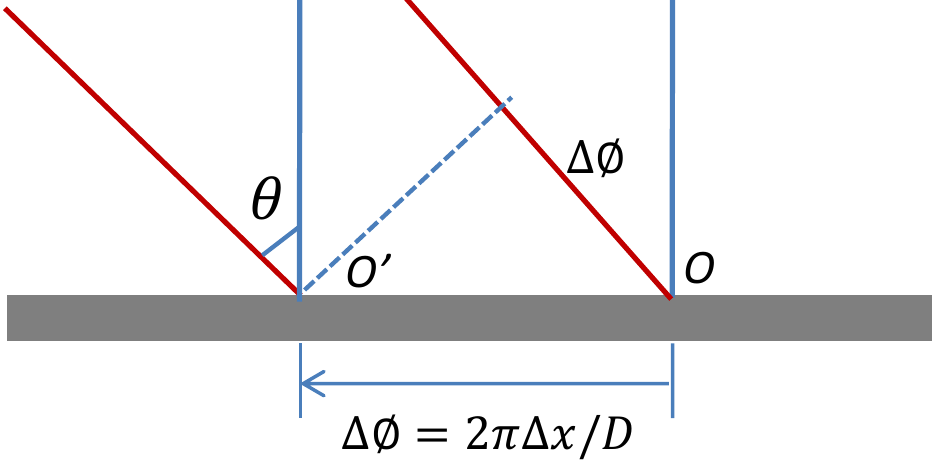}\label{fig:fig6b}}
	   	  	  \subfigure[]{\includegraphics[height=4.8cm,width=1\linewidth]{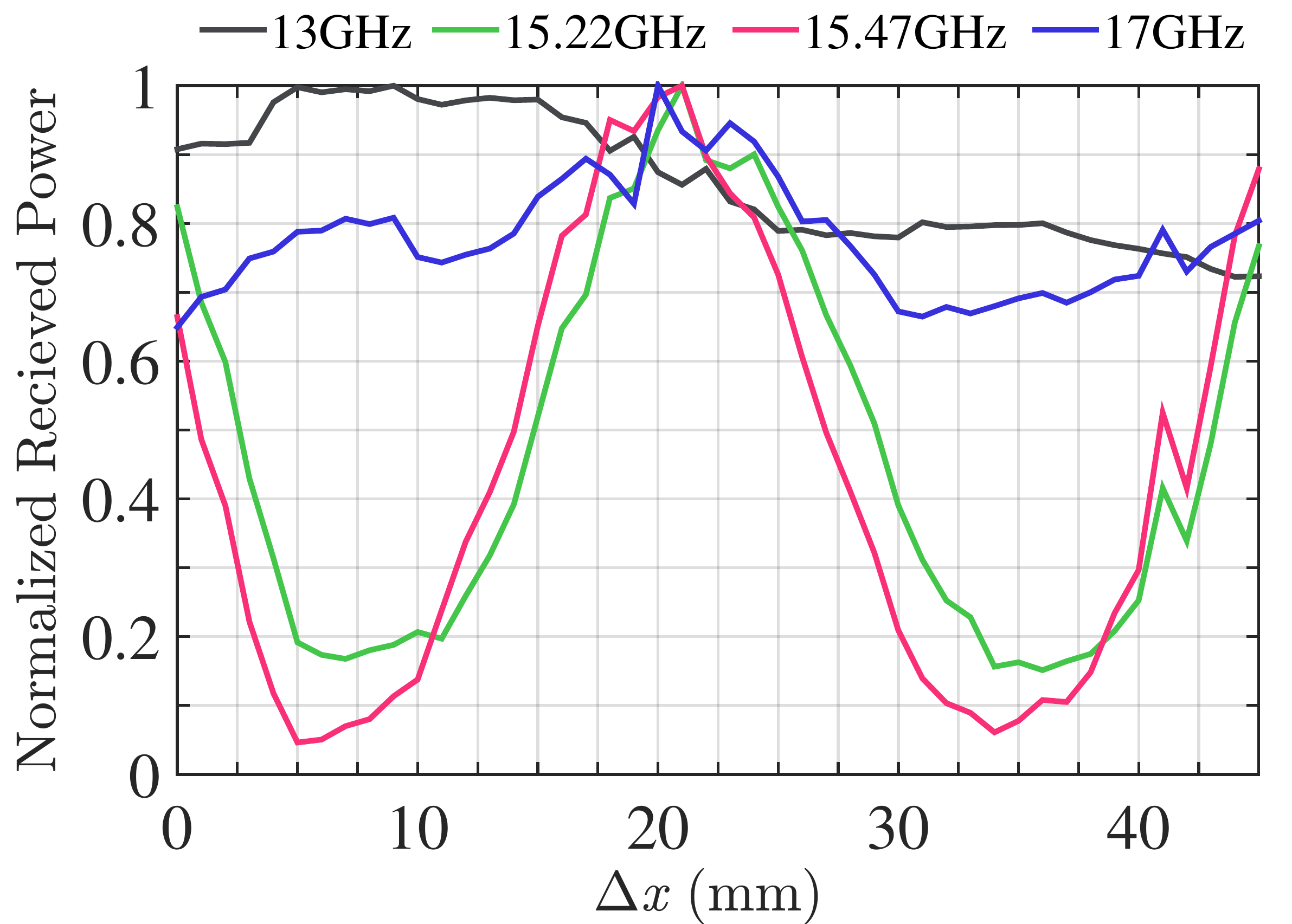}\label{fig:fig6c}}
	   			\end{minipage} \par	
	\caption{ 
			(a) Experimental setup. Two transmitting antennas fed via a power divider illuminate the metasurface normally and at $45^\circ $. A receiving antenna is placed at $45^\circ $ to measure the total reflected power. Due to the periodicity of the metasurface,  continuously-changing phase difference between the two beams can be emulated by moving the metasurface horizontally along the impedance variation direction. Two pieces of  absorbing foam are put on both sides, ensuring that the effective exposure area of the metasurface remains fixed when the surface is shifted. (b) The reference point $O$ is the intersection point of the $ 0^\circ $ and $ 45^\circ $ beams on the metasurface when the phase difference is $0$. The phase difference at a distance  $\Delta x$ from the reference point $O$ is $\Delta\phi=2\pi \Delta x/D$, which is linearly varying as a function of the horizontal distance $\Delta x$. (c) The normalized received power for different metasurface positions at 13, 15.22, 15.47, and 17 GHz. 
			}  
			\end{figure*}	
        
To experimentally verify the phase-controlled reflection by the metasurface, in the last measurement shown in Fig.~\ref{fig:fig6a}, two transmitting antennas fed via a power divider illuminate the metasurface normally and at $45^\circ$. A receiving antenna is placed at the $45^\circ$ angle to measure the total power reflected by the metasurface under two simultaneous illuminations. To avoid severe insertion loss caused by the use of a phase shifter in one branch, which may increase the amplitude inequality between two beams, we mimic the phase-difference-tuning process by moving the metasurface along the $x$ direction. As seen in Fig.~\ref{fig:fig6b}, the phase difference between the two beams is linearly varying when we change the horizontal  position of the metasurface. Therefore, this shift is equivalent to a phase change between the two beams. To ensure the effectively-illuminated area of the metasurface to remain stable during the moving process, we put two pieces of absorbing foam on top of both sides of the sample. The total received power, normalized by the maximum power of reflected wave is changing with varying the distance $\Delta x$. As is seen in Fig.~\ref{fig:fig6c}, the modulation depths reach $0.15$ and $0.04$ at $15.22$~GHz and $15.47$~GHz, respectively. This result indicates that coherent enhancement and cancellation near the design frequency can be achieved by tuning the phase difference of the two incident beams. The period of the modulation is about $29$~mm, almost equal to the period of the metasurface, which validates the theoretical analysis. However, at the frequency far from the designed one, for instance at $13$~GHz and $17$~GHz, the coherent phenomenon becomes much weaker, as is seen in Fig.~\ref{fig:fig6c}, due to a mismatch of the main reflection angles and the reflection amplitudes of the normally and obliquely incident waves.

\section{Discussion}	
We have demonstrated coherent perfect absorption of two beams incident at arbitrary angles. It has been found that this effect is possible for relative beam amplitudes within a certain range using a gradient passive planar structures. When these two incidences change into out-of-phase state, reflections at all three propagating channels come out. To realize coherent control of reflection with single direction, the other parasitic reflections can be suppressed by introducing unidirectional evanescent modes excitation. To realize a larger reflection for out-of-phase scenario, we use an optimization algorithm to search for an optimum solution of grid impedance profile and substrate thickness, which is powerful when many degrees of freedom are required in multi-channel metasurface design. In the other design methodologies such as non-local metasurface \cite{D2017From} and plasmonic grating \cite{2012Measurement,Janus_scattering_ZJ_WANG}, where the interference between all the elements of a unit cell are important for the device performance, a brute-force optimization process in full-wave simulations is required, which is time consuming and even cannot work when multiple input beams and multi-functionalities for multiple channels are involved. Compared with them, our approach is much more robust and efficient due to a rigorous theoretical analysis, particularly by introducing unidirectional evanescent mode in the scattered field to eliminate parasitic reflections. Moreover, the angle-dependence of the impedance of substrate is also considered in our algorithm, which is vital in metasurface design for multiple-angle incidence scenarios \cite{zhang2020controlling,yuan2021control}.

We have  realized a gradient metasurface with angular-asymmetric coherent perfect absorption and reflection functionalities. The concept of wave control via evanescent harmonics engineering and independent control of the electromagnetic response for multiple illuminations can be applied for engineering multi-functional wave processes. Metasurface-based designs are attractive in practical applications. For example, by placing a planar structure on a metal-grounded dielectric layer, the velocity or position of the object can be detected by monitoring the total reflection of such a object under two coherent illuminations. Additionally, we hope that this work can find promising applications in phased-array antennas, one-side detection and sensing, and optical switches with low insertion loss.
	
\section{Methods}
\textbf{Design and modeling of the metasurface}

The prototype presented in this work was designed for operation at $15.22$~GHz. The grid impedance is discretized into 6 sub-cells, and each sub-cell is divided into 4 equal sub-sub-cells. The effective grid impedance of each sub-sub-cell is retrieved from simulated reflection coefficient ($S_{11}$) through the transmission-line method approach (see the Supplementary Material\cite{See}). Numerical simulations are carried out using a frequency-domain solver, implemented by CST MWS. Excitations propagating along the $z$-direction from port 1 with the electric field along the $y$-direction and the magnetic field along the $x$-direction are used in the simulations to obtain the $S_{11}$ parameter. The dimensions of all the elements in the unit cells are designed and optimized one by one to fit the theoretically found required surface impedance.

Once the dimensions of all the elements in the unit cells are found, we perform numerical simulations of the unit cell in CST MWS for the normal and  $ 45^\circ $ incidences. The simulation domain of the complete unit cell was $D \times D_y \times  D$ (along the $x, y$, and $z$ directions), the unit cell boundary condition and the Floquet port were set. The scattered fields for the normal and $ 45^\circ $ incidences were calculated by subtracting the incident waves from the total fields. Finally, the total scattered fields when the metasurface is illuminated by two waves silmutaneously were obtained by adding the scattered field of each single beam with different phase differences. 

\textbf{Realization and measurement}

The ITO pattern of the metasurface was manufactured using the picosecond laser micromachining technology on a 0.175-mm-thick ITO/PET film. The sample comprises 10 unit cells along the $x$ axis and 66 unit cells along the $y$ axis [Fig.~\ref{fig:fig5b}] and has the size of $14.15\lambda \times 10.04\lambda = 278.9~mm \times 198~mm$. The ITO/PET film was adhered to a 4.95-mm-thick F4BTM substrate with $\epsilon= 5.8(1-j0.01)$ backed by a copper ground plane.

The operation of the designed metasurface was tested using a NRL-arc setup [Fig. ~\ref{fig:fig5b}]. In the experiment, two double-ridged horn antennas with 17 dBi gain at $15.22$~GHz are connected to a vector network analyzer as the transmitter and receiver. The metasurface was located at a distance of 2 m (about 101$\lambda$) from both the transmitting and receiving antennas where the radiation from the antenna can be approximated as a plane wave. The antennas are moved along the scanning track to measure the reflection towards different angles. Time gating is employed to filter out all the multiple scattering noise signals received by the antenna \cite{See}. 

\section{DATA AVAILABILITY}
The data that support the findings of this study are available from the corresponding authors upon reasonable request.

\bibliography{Reference} 
\bibliographystyle{naturemag}

\section{Acknowledgements}
The authors are grateful to Dr. Viktar S. Asadchy for useful discussions. S.M.Z. acknowledges support from China Scholarship Council. This research was also supported by the Natural Science Foundation of Zhejiang Province(LY22F010001), the Natural Science Foundation of China (61701268), and the Fundamental Research Funds for the Provincial Universities of Zhejiang.

\section{Author contributions}
S.M.Z. and X.C.W. conceived the study. S.M.Z. performed the numerical calculations, and designed the samples. S.M.Z. conducted the experiment. S.M.Z., X.C.W., and S.A.T. wrote the paper. S.A.T. supervised the project. All authors contributed to scientific discussions and  editing  the manuscript.
\section{Competing interests}
The authors declare no competing interests.

\section{Additional information}
\textbf{Supplementary information} The online version contains supplementary material available at https:xxxx.

{\bf Correspondence} and requests for materials should be addressed to Shuomin Zhong or Xuchen Wang.

\end{document}